\documentclass{JHEP3} 

\usepackage{epsfig,multicol}

\newcommand{\ds}{{\sffamily DarkSUSY}}
\def\msun{M_{\odot}{\ }}
\newcommand{\code}[1]{{\tt #1}}

\newcommand\fverb{\setbox\pippobox=\hbox\bgroup\verb}
\newcommand\fverbdo{\egroup\medskip\noindent%
			\fbox{\unhbox\pippobox}\ }
\newcommand\fverbit{\egroup\item[\fbox{\unhbox\pippobox}]}
\newbox\pippobox
\newcommand{\beq}{\begin{equation}}
\newcommand{\eeq}{\end{equation}}
\def\sss{\scriptscriptstyle}

\title{The Role of Antimatter Searches in the\\ Hunt for Supersymmetric Dark Matter}
\author{Stefano Profumo and Piero Ullio\\
	SISSA/ISAS, via Beirut 2-4, 34013 Trieste, Italy\\
	and Istituto Nazionale di Fisica Nucleare, Sez. di Trieste, 34014 Trieste, Italy\\
	E-mail: \email{profumo@sissa.it}, \email{ullio@sissa.it}}

\preprint{SISSA-37/2004/EP}

\abstract{We analyze the antimatter yield of supersymmetric (SUSY) models with large neutralino annihilation cross sections. We introduce three benchmark scenarios, respectively featuring bino, wino and higgsino-like lightest neutralinos, and we study in detail the resulting antimatter spectral features. We carry out a systematic and transparent comparison between current and future prospects for direct detection, neutrino telescopes and antimatter searches. We demonstrate that often, in the models we consider, antimatter searches are the only detection channel which already constrains the SUSY parameter space. Particularly large antiprotons fluxes are expected for wino-like lightest neutralinos, while significant antideuteron fluxes result from resonantly annihilating binos. We introduce a simple and general recipe which allows to assess the visibility of a given SUSY model at future antimatter search facilities. We provide evidence that upcoming space-based experiments, like PAMELA or AMS, are going to be, in many cases, the unique open road towards dark matter discovery.}


\begin{document} 

\section{Introduction}

Weakly interacting massive particles (WIMPs) are among the leading candidates for cold dark matter in the Universe: on one side, the idea is appealing since dark matter appears just as another of the thermal leftovers from the early Universe; at the same time, from the particle physics point of view, numerous models naturally embed such kind of particles. The search for WIMPs has therefore been one of the hot topics in experimental physics over the last decade or so, and the field is now acquiring new perspectives with next-generation experiments being at the research and development stage, or, in some cases, even closer to the data taking phase.  The target is either to detect WIMPs directly, measuring the recoil energy from WIMPs elastic scattering on nuclei, or to search for the yields of WIMP pair annihilations, such as neutrinos from the Earth/Sun or cosmic rays from annihilations in the galactic halo. Complementarity between direct detection and indirect detection methods has been repeatedly stressed (for reviews on this topic, see, e.g., \cite{jkg,larsrev}). Indeed, the idea of WIMP dark matter refers to a broad framework rather than to any specific model; in such a framework, an overall (weak) coupling of dark matter to ordinary matter is guaranteed, but details on how such coupling is realized may be sensibly different from one model to another: One of the most pressing issues from an experimentalist perspective, the question regarding how competitive a given detection technique is compared to the others, can be addressed just on model dependent grounds.

The most popular WIMP dark matter candidate is the lightest neutralino in supersymmetric extensions of the standard  model of particle physics. The idea of supersymmetry has strong theoretical motivations, but it has, unfortunately, no experimental confirmation so far. Recipes on how to introduce a supersymmetric model (and hence the corresponding dark matter candidate) have rather weak constraints; to some extent, this freedom is even too large, so that, for practical purposes, it is customary to focus on minimal models, i.e. models with minimal particle content and a small number of free parameters. The benchmark scenario most often implemented is the mSUGRA setup, i.e. the supergravity inspired model with the smallest possible number of free parameters, namely only four parameters and one sign, partly defined at the grand unification (GUT) scale and partly at the low energy (electroweak) scale.  Referring, within this framework, to models with a thermal relic density (in a standard setup) matching the cold dark matter term as deduced from latest high precision cosmological measurements \cite{Spergel:2003cb}, prospects for direct detection look more promising than those for indirect detection methods (see, e.g., the recent analysis in ref.~\cite{esu}). To some extent, a similar conclusion is reached also for thermal relic models in the context of minimal supersymmetric standard model (MSSM) setups defined by few universal parameter chosen at a low energy scale (such as the seven parameter model implemented, e.g., in ref.~\cite{someds}). There is often, in the literature, the tendency to extrapolate on these results, and assume that, as a generic trend for supersymmetric dark matter candidates, direct detection should have top priority.

We give here a few counter examples, still in the context of minimal supersymmetric setups, which allow for a clearer presentation of results and a more transparent comparison between different signals (for related studies within various supersymmetry breaking scenarios see, e.g., \cite{dmother}). We will enlighten, in particular, the role of indirect detection through the search for exotic antimatter components in cosmic rays.
This technique was proposed soon after the formulation of the idea of WIMP dark matter, the links to the framework definition being evident: Pair annihilation is the mechanism which sets the thermal relic abundance of WIMPs; although the density of WIMPs in dark matter halos today is much smaller than in the early Universe environment, there is still a finite probability for WIMPs in the Galactic halo to annihilate in pairs. In these annihilations the same amount of matter and antimatter is produced; while the matter component is likely to be very subdominant compared to standard astrophysical sources, there seems to be no standard primary source of antimatter, with the bulk of the (scarce) antimatter component in cosmic rays which is likely to be of secondary origin, i.e. generated in the interaction of primary cosmic rays (mainly protons) with the interstellar medium (mainly hydrogen and helium). The goal is then to identify, through their peculiar spectral features, the WIMP-induced antimatter fluxes, or at least to exclude those dark matter candidates which would overproduce antimatter compared to the relatively low background term.

Implementing models for the dark matter halo with self-consistent density and velocity distributions, we will produce fully consistent comparisons of indirect detection versus direct detection, examining both current exclusion limits and future detection prospects. Our analysis will be focussed on a few sample particle physics scenarios, but to facilitate the extension to alternative frameworks, we will introduce a simple and general recipe to address detectability of a WIMP model in future antimatter cosmic ray measurements.

The outline of the paper is as follows. After introducing the particle physics setup in sec.~\ref{sec:parphys}, we provide, in sec.~\ref{sec:SUSYantimatter}, details on the computation of antimatter fluxes from neutralino annihilations, and describe the relevant spectral features. Sec.~\ref{sec:curlim} is devoted to a comparison among current exclusion limits from different search strategies, while in sec.~\ref{sec:futlim} we present a novel technique for the evaluation of future discrimination capabilities at antimatter searches, and confront the projected experimental sensitivities at direct and indirect future dark matter detection experiments. The question of the dependence of antimatter yields on the halo model is treated in sec.~\ref{sec:hmodels}. Finally, sec.~\ref{sec:conclusions} concludes. 

\section{The particle physics framework}\label{sec:parphys}

We suppose that our cold dark matter candidate is the lightest neutralino $\widetilde{\chi}^0_i$ as defined in a MSSM setup, i.e. the lightest mass eigenstate from the superposition of the two neutral gaugino and the two neutral Higgsino fields:
\begin{equation}
  \widetilde{\chi}^0_1 = 
  N_{11} \widetilde{B} + N_{12} \widetilde{W}^3 + 
  N_{13} \widetilde{H}^0_1 + N_{14} \widetilde{H}^0_2\;.
\end{equation}
The {\em bino}, {\em wino} and {\em higgsino} fractions, which we define, respectively, as $|N_{11}|$, $|N_{12}|$, and$\sqrt{N_{13}^2+N_{14}^2}$, are obtained by diagonalizing the neutralino mass matrix, and depend mainly on the relative hierarchy between the bino, the wino and the higgsino mass parameters, i.e., respectively, $M_1$, $M_2$ and $\mu$. The WIMP relic abundance and detection prospects are mainly set by the lightest neutralino mass and composition, but depend in part also on the structure of the Higgs and sfermion sectors of the theory. We will select a few specific configurations, and to clarify how this choice is made, we need to briefly review which role is played by the relic density and by other constraints on the model.

\subsection{Relic abundance, accelerator constraints and WIMP models}

Let's start our discussion in the context of the popular {\em minimal Supergravity} (mSUGRA) framework~\cite{msugra}. In this scenario, the lightest neutralino is  {\em bino}-like over most of the parameter space, as a consequence of gaugino soft SUSY breaking mass unification at the GUT scale; the only exception is a rather small parameter space region, often dubbed {\em focus-point} or {\em Hyperbolic Branch} region~\cite{focus}, close to the region where electroweak symmetry breaking (EWSB) conditions can no longer be fulfilled, where a low value of the $\mu$ parameter is achieved through large values of the scalar mass parameter $m_0$. 

Binos have rather low pair annihilation rates (mainly into fermion-antifermion pairs, much smaller than higgsinos or winos  annihilation rates into weak bosons), getting lower and lower for increasing bino masses. As the WIMP relic abundance scales with the inverse of the annihilation rate (the higher the rate, the longer the WIMP is kept in thermal equilibrium, the smaller the WIMP density at thermal decoupling and hence today), it follows that in most of the mSUGRA parameter space, the lightest neutralino relic abundance is larger than the upper WMAP bound on $\Omega_{\rm CDM} h^2$.
One should then restrict to light binos, in the so-called {\em bulk} region at small values of $m_0$ and of the soft supersymmetry breaking fermionic mass parameter $m_{1/2}$ (the one through which $M_1$, as well as $M_2 \simeq 2 M_1$, is fixed at the low energy scale); however, the bulk region is almost entirely excluded  by accelerator constrains since, at low values of $\tan\beta$, the ratio of the vacuum expectation values of the two neutral components of the SU(2) Higgs doublets, the LEP constraint on the lightest $CP$-even Higgs boson mass $m_h$ is violated, while, at large $\tan\beta$, stringent bounds stem from the inclusive $b\rightarrow s\gamma$ branching ratio
(for reference, throughout this analysis we will assume as a bound on $m_h$ the standard model Higgs bound of 114.1~GeV~\cite{SMHiggs}, and an allowed range $2.0\times10^{-4}\leq\mathrm{BR}[B\rightarrow X_s\,\gamma] \leq4.6\times10^{-4}$, obtained adding a theoretical uncertainty of $\pm0.5\times10^{-4}$ to the
experimental value quoted by the Particle Data Group 2002~\cite{pdg02}, with an estimate of this process including the complete next-to-leading 
order (NLO) correction for the standard model contribution~\cite{bsgsm,bsgmagic} and the dominant NLO corrections for the SUSY term~\cite{bsgtan,bsgsusy}).

The loopholes for bino dark matter in this scenario are found in two regimes: {\sl1)} the tiny coannihilation branches, i.e. stripes of the parameter space where the lightest neutralino is just slightly lighter than the next-to-lightest  SUSY particle, i.e., in this scenario, a stau at low $m_0$~\cite{Ellisstau1,Ellisstau2} or a stop for large trilinear couplings~\cite{Ellisstop}, which drives the thermal decoupling in the early Universe; {\sl 2)} the {\em funnel} region, where annihilations into fermions through $s$-channel exchanges of the $CP$-odd $A$ (or of the heaviest $CP$-even $H^0_1$) neutral Higgs boson are enhanced by the resonance condition $m_\chi \simeq m_A / 2$ (or $m_\chi \simeq m_{H^0_1} / 2$). In both regimes, and in the lower neutralino mass end, direct detection rates are well below current sensitivities, but above the projected sensitivity of future large-mass experiments. Unfortunately this does not follow from crossing symmetry, i.e. from some diagrammatic structure giving large contributions to both the annihilation rate and the scattering rate on light quarks. The scalar superpartners of light quarks are very heavy, hence scattering rates are dominated by $t$-channel exchanges of two $CP$-even Higgs; these contributions are guaranteed to be large in this framework, as the higgsino fraction is never forced to zero (the coupling of neutralinos to a Higgs boson scales with the mixing between bino and higgsino components, or wino and higgsino terms) and the structure of the scalar sector is very constrained (for negative $\mu$ however, the two contributions  have negative interference and the scattering cross section may become small). 

From the point of view of indirect dark matter detection, the regime with bino-like neutralinos in the coannihilation tails is twice disfavored: first there is the S-wave suppression of the cross section for the annihilation of non-relativistic Majorana fermions into light fermions (at non-zero temperature, in the early Universe, the annihilation cross section is dominated by the P-wave); at the same time, the effective annihilation cross section $\langle\sigma_{\rm eff}v\rangle$ at freeze-out is dominated by coannihilating partner contributions, with the actual neutralino annihilation at $T=0$ being much smaller. The funnel regime is instead more interesting, since resonant annihilations would still be effective at $T=0$, giving large values for  $\langle\sigma_{\rm eff}v\rangle(T=0)$. Actually, most often, the enhancement effect is not at full strength, because the neutralino thermal relic abundance of models matching too closely the resonance condition tends to be smaller than needed: the choice falls then on models slightly off-resonance, again with an effect that is less dramatic for the thermally averaged term in the early Universe than for the annihilation cross section at zero temperature in today's structures.

A similar feature emerges in the focus point region, or in MSSM models defined at the weak scale keeping $\mu$ and $M_1$ as free parameters, but still, say, with some common mass scale in the squark sector, which again is forced to be heavy by accelerator constraints. Requiring the thermal relic density to lie within the WMAP selected range, is equivalent, at a given neutralino mass, to fix the mixing between bino and higgsino components; except for very heavy neutralinos in the TeV range, where detection with any method looks in any case problematic, the mixing is large, and the annihilation rate into weak gauge bosons (which is not S-wave suppressed) is never at full strength, with again, to some extent, low temperature effects playing against indirect detection. The large mixing enhances instead the direct detection rate, but again this is not an effect of crossing symmetry, i.e. a feature which follows automatically from the definition of thermal WIMP.

From this discussion it emerges that the hierarchy between direct and indirect detection may be inverted, in supergravity inspired models, relaxing some of the so-called {\em universality assumptions}, for instance with more complex scalar sectors~\cite{scalarext}. We choose here, instead, to restrict to minimal scenarios, and to relax, instead, the constraint coming from the computation of the thermal relic density. We will include in our discussion models with low relic densities, or, more precisely, entertain the possibility that the WIMP relic abundance is enhanced compared to the standard scenario. There are numerous possibilities on how such enhancement may take place, ranging from cosmological scenarios with faster expansion rates at the time of decoupling, such as in cosmologies with a quintessence energy density term dominating at the WIMP freeze-out temperature~\cite{quint,Profumo:2003hq,Profumo:2004ex}, anisotropic Universes with effective shear density terms~\cite{shear,Profumo:2004ex} or in scalar-tensor theories \cite{Catena:2004ba}, to particle physics models with non-thermal sources of WIMP dark matter, such as neutralino productions from gravitinos or moduli decays~\cite{non-therm}.

\subsection{The three benchmark scenarios}\label{sec:SUSYmodels}

We focus on three minimal SUSY frameworks with the lightest neutralino being respectively bino, wino and higgsino like, each of them in large purity configurations. All models are soundly motivated from the high energy physics setup point of view. The neutralino mass ranges within values allowed by all accelerator direct and indirect bounds. 

\underline {\large{\em Binos}: the {\bf Funnel} model.}\,\, We consider the mSUGRA scenario, and select models within the funnel region, along a section of the $m_{1/2}$-$m_0$ parameter space in which the lightest neutralino mass {\em exactly matches} the resonance condition with the $CP$-odd Higgs boson mass, i.e. $2\cdot m_\chi = m_A$.  In mSUGRA, the tree level value of $m_A$ is fixed by $\tan\beta$ and by the sign of $\mu$ through EWSB Conditions; taking into account that we need the lightest neutralino to be the LSP, the resonance is realized only at $\tan\beta\gtrsim 35$ in case sgn$\mu<0$, or at $\tan\beta\gtrsim50$ if sgn$\mu>0$. Since for $\mu<0$ and large $\tan\beta$ the SUSY contributions to $BR(b\rightarrow s\gamma)$ are large, and rule out models up to fairly heavy neutralinos, we consider an example with positive $\mu$,  $\tan\beta=55$ and trilinear coupling $A_0=0$ (the latter is fixed for definiteness, as it does not play much of a role). The resonance curve in this configuration starts at a minimum value of $m_{1/2}$ (and therefore $m_\chi$) below which the lightest stau is lighter than the lightest neutralino\footnote{In this low mass region, stau coannihilation effects are also present, and are taken into account in the computation of the relic abundance; they are, however, not relevant in the context of dark matter searches.}, and ends at at maximum $m_{1/2}$ corresponding to the model with neutralino relic abundance exceeding the CDM upper limit. The bino purity of these models is anywhere significantly large, always above 99.6\% (see tab.~\ref{tab:modelsneut}). We will hereafter indicate this set of configurations as the {\bf Funnel} model.


\TABLE[t]{\begin{tabular}{|c|c|c|c|c|}
\hline
Model & $M_{1/2}$, $m_{3/2}$, $M_3$ & $\tan\beta$ & sgn($\mu$) & Defining Condition \\
\hline
\hline
{\bf Funnel} & $700 \div 1450$ & 55 & $>0$ & $2\ m_\chi\simeq m_A$ \\
{\bf AMSB} & $23 \div 231$ & 50 & $>0$ & $m_0=1500$ GeV\\
{\bf NUGM} & $879 \div 1096$ & 50 & $>0$ & $M_1/M_3=10$, $M_2/M_3=2$, $\widetilde{H}=99.8$\%\\
\hline
\end{tabular}
\caption{The three SUSY models under consideration. Models {\bf Funnel} and {\bf NUGM} have mSUGRA-like boundary conditions at the high energy scale, while model {\bf AMSB} features minimal Anomaly Mediated SUSY Breaking. $M_{1/2}$ indicates the value (in GeV) of the common gaugino mass at the grand unification scale $M_{\rm\sss GUT}$ in model {\bf Funnel} (second column); in model {\bf AMSB} the neutralino mass is instead driven by the value $m_{3/2}$ of the gravitino mass (in TeV, second column); finally, model {\bf NUGM} ({\em non-universal gaugino masses}) has the gaugino non-universality pattern of the {\bf 200} representation of the symmetric product of two $SU(5)$ adjoints, and the high energy parameter we use is the gluino mass $M_3$ (in GeV, second column). For each model, the value of the common scalar SUSY breaking mass $m_0$ is dictated by the {\em Defining Condition}, which is indicated in the last column of the Table. The scalar trilinear coupling $A_0=0$ for all models.}\label{tab:models}}

\TABLE[t]{\begin{tabular}{|c|c|c|c|}
\hline
Model & Bino fraction & Wino fraction & Higgsino fraction \\
\hline
\hline
{\bf Funnel} & $>99.6$\% & $<0.05$\% & $<0.5$\%\\
{\bf AMSB} & $<0.02$\% & $>98$\% for $m_\chi>100$ GeV &  $<2$\% for $m_\chi>100$ GeV\\
{\bf NUGM} & $<0.01$\% & $<0.2$\% & $99.8$\%\\
\hline
\end{tabular}
\caption{The lightest neutralino composition for the three SUSY models of Tab~\ref{tab:models}, in terms of the bino, wino and higgsino fractions.}\label{tab:modelsneut}}


\underline {\large {\em Winos}: the {\bf AMSB} model.}\,\, In the so-called minimal Anomaly Mediated SUSY Breaking (mAMSB) scenario \cite{Randall:1998uk,Giudice:1998xp,Gherghetta:1999sw,Feng:1999hg} the gaugino masses are proportional, through the gravitino mass $m_{3/2}$, to the beta functions of the $SU(3)$, $SU(2)$ and $U(1)$ gauge groups, therefore drastically affecting the soft breaking gaugino mass hierarchies at low energies, as compared to a supergravity and grand-unification inspired scenario: the lightest neutralino is in fact always {\em wino}-like, with a remarkably large purity. This feature induces  a nice property of the mAMSB scenario, i.e. the homogeneity of observables over its parameter space: except for accidental cancellations, the relic abundance and the rates for indirect detection are sensitive to the value of the LSP mass, but have a rather mild dependence on other specific features of the low energy structure of the theory. We decided to resort to a rather large value of the common scalar mass parameter $m_0=1500$ GeV, and of $\tan\beta=50$. The sign of $\mu$ has been assumed to be positive. The range of the gravitino mass (and therefore of the lightest neutralino) is bounded from below by the null-results of chargino searches at LEP~\cite{chargino}\footnote{In the case of mAMSB, the LEP bound on the chargino mass is somewhat weakened due to the very small splitting between the lightest neutralino and the lightest chargino. We will nonetheless restrict to values larger than 100 GeV.}, and from above by the fact that one enters in the region where EWSB can no longer be fulfilled, giving raise to an unphysical Higgs sector. We label the configuration defined by this choice of parameters, which is summarized in tab.~\ref{tab:models}, as {\bf AMSB} model; details on the composition of the lightest neutralino are given in tab.~\ref{tab:modelsneut}.

\underline {\large {\em Higgsinos}: the {\bf NUGM} model.}\,\, The assumption that gaugino masses unify at the GUT scale $M_{\sss\rm GUT}$ may be relaxed in a number of ways. In the context of supergravity, non-vanishing gaugino masses are generated through the SUSY breaking vev of the auxiliary component of the lowest order non-renormalizable term in the gauge kinetic function. The SUSY breaking vevs, in order to preserve gauge invariance, must lie in a representation belonging to the symmetric product of two adjoints of the underlying unified gauge group. In the case of $SU(5)$ GUTs, one has
\begin{equation}
\left({\bf 24}\times{\bf 24}\right)_{\rm symm}={\bf 1}\oplus{\bf 24} \oplus{\bf 75}\oplus {\bf 200}.
\end{equation}
Universal gaugino masses are generated {\em only} if the SUSY breaking field lies in the singlet representation {\bf 1}. In all other cases, particular ratios between gaugino masses will hold at the GUT scale \cite{Ellis:1985jn,Drees:1985bx}. In particular, two of the four representations yield a {\em higgsino-like} LSP, namely the {\bf 75} and the {\bf 200}. We focus here on the {\bf 200}, because in the case of the {\bf 75} successful EWSB forces the parameter space to rather narrow regions, and to low values of $\tan\beta$ \cite{Chattopadhyay:2003yk}. The parameter space in the corresponding minimal scenario is analogous to the mSUGRA case, with a gaugino mass parameter, which we take to be the gluino mass $M_3$, and a common scalar mass parameter $m_0$ (plus again $\tan\beta$, $A_0$ and sgn($\mu$)). 
Since we want to study the case of a pure higgsino, we solve the parameter space degeneracy by requiring a fixed higgsino content of the lightest neutralino, $\widetilde{H}=99.8$\%. This specific value has been chosen in order to maximize the neutralino mass range. $m_\chi$ turns out to decrease at growing $m_0$, along the iso-higgsino content lines; analogously to our previous choices, we fix $\tan\beta=50$, $A_0=0$ and sgn$\mu>0$, and let $m_0$ vary from 0 to the largest value compatible with EWSB, achieving a rather wide neutralino mass range. Details on the model are provided in the bottom line of tab.~\ref{tab:models}; this configuration will be indicated from now on as {\bf NUGM} (non-universal gaugino mass) model.

\FIGURE[t]{\epsfig{file=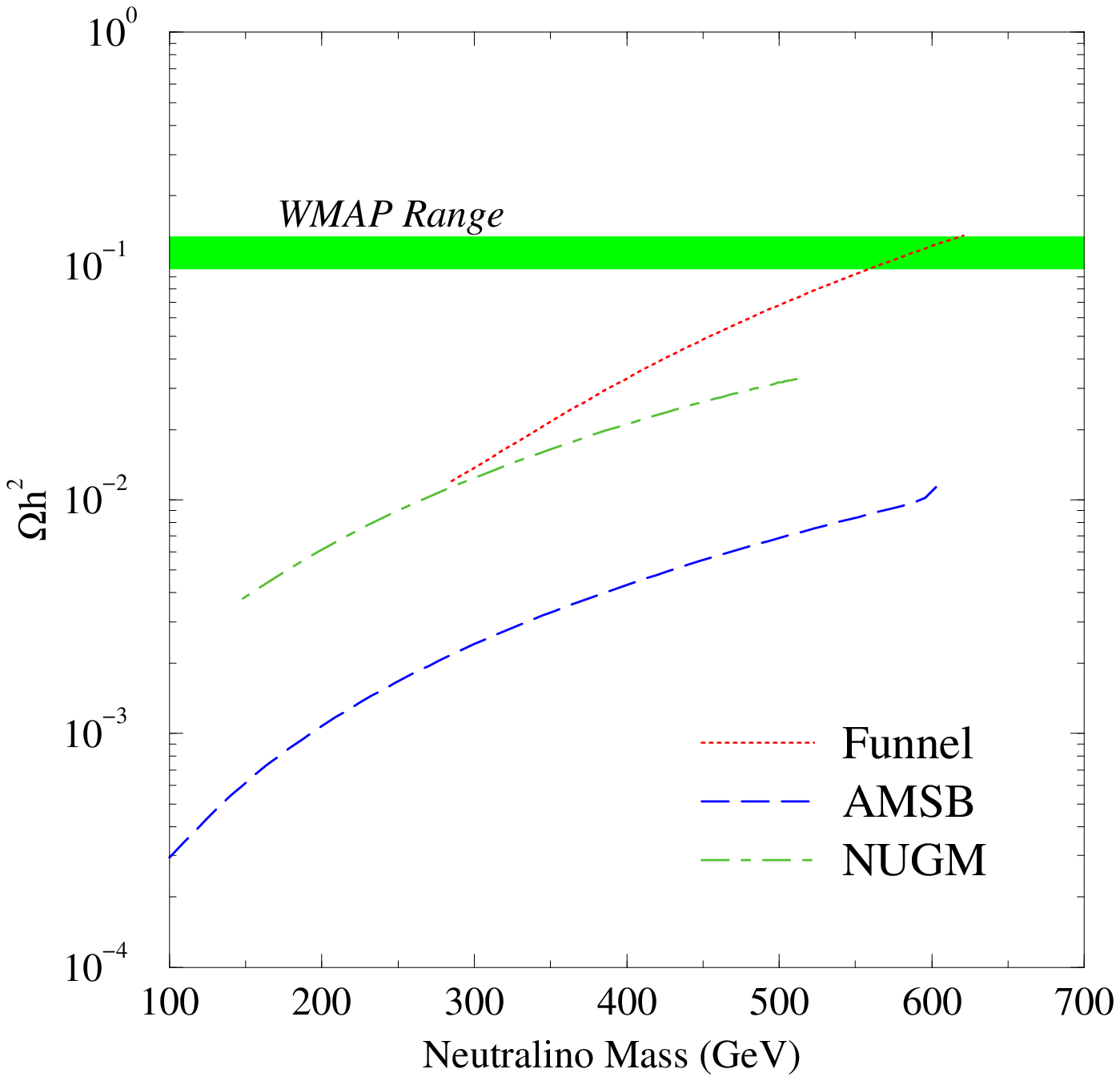,width=7.1cm}\qquad
\epsfig{file=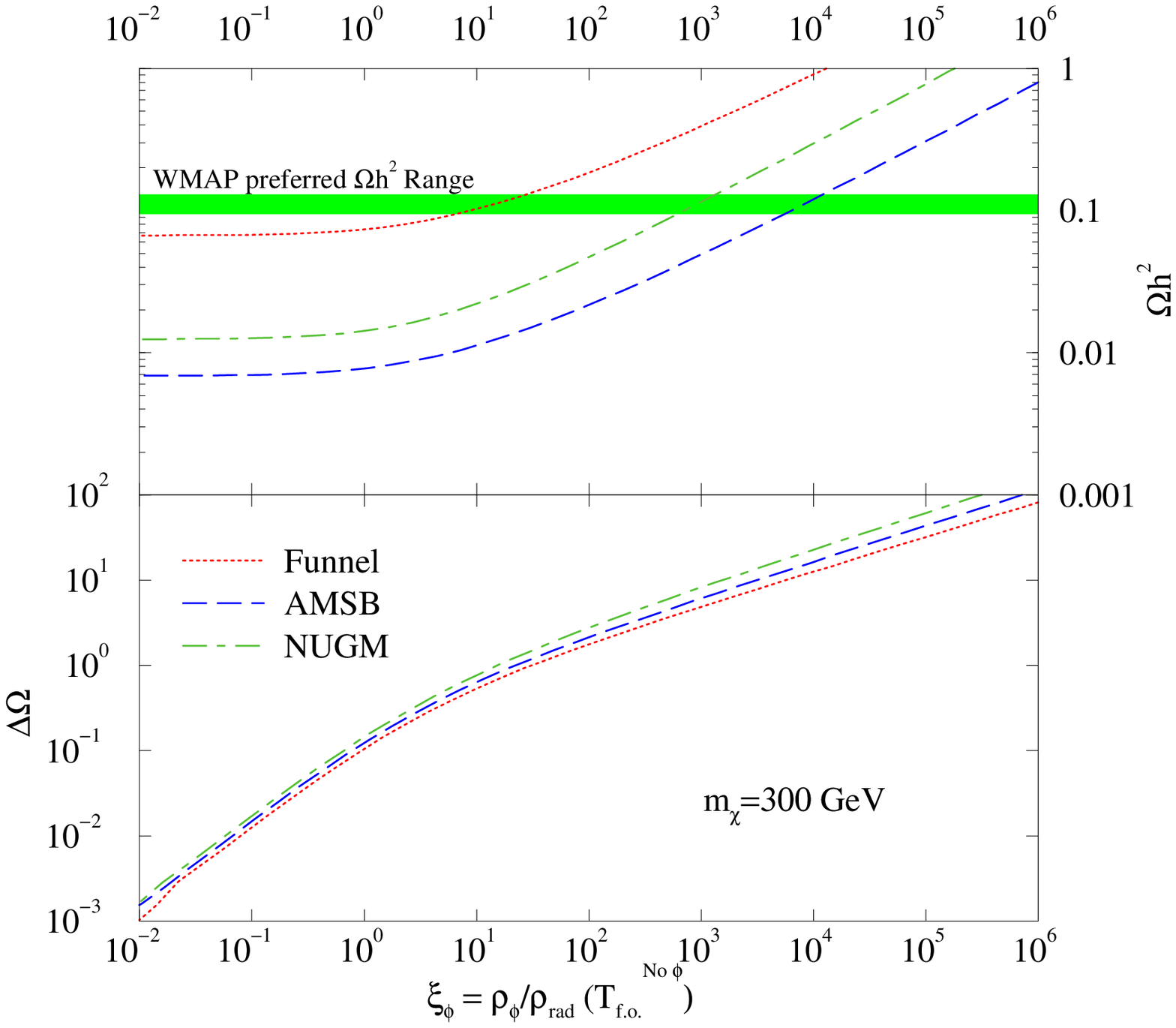,width=7.25cm} 
\caption{In the figure to the left we show the neutralino relic abundance, in the standard cosmological scenario, for the three models described in the text (see Tab.~\ref{tab:models}), respectively featuring a resonantly annihilating bino ({\bf Funnel}), a wino ({\bf AMSB}) and a higgsino ({\bf NUGM}) lightest neutralino, as a function of the neutralino mass. The green band indicates the preferred WMAP range \cite{Spergel:2003cb}. The right panel shows instead the relic density enhancement within the quintessential scenario of ref.~\cite{Ferreira:1997hj}, i.e. with an exponential potential, with the exponent $\lambda=3.5$, for the three benchmark models. The upper part shows the relic abundance, while the lower part the relative enhancement $\Delta\Omega=(\Omega_Q-\Omega)/\Omega$, both as a function of the parameter $\xi_\phi$, the ratio of the quintessential and the radiation energy densities at the neutralino freeze-out temperature without quintessence.}
\label{fig:oh2}}

For all the three benchmarks scenarios introduced above, soft breaking parameters, gauge and Yukawa couplings are evolved down to the weak scale with the ISASUGRA RGE code as given in version 7.67 of the ISAJET software package~\cite{isajet} (introducing some minor changes, such as, the conversion of the ISASUGRA code to double precision to improve on its stability; for more details on this and other technical points on the code implementation and on the interface with the \ds\ package, see the discussion in~\cite{coann}). Weak scale spectra are interfaced into the \ds\ computer code \cite{ds}, which is then used for computations of the relic density and of direct and indirect detection rates. Regarding, in particular, the relic abundance calculations, we rely on a code which provides a high-precision fully-numerical treatment of resonances and coannihilation effects, here necessarily needed to properly include the resonance on the $A$ Higgs boson for 
the {\bf Funnel} model and chargino coannihilations for the {\bf AMSB} and {\bf NUGM} models (for both pure winos and higgsinos, the lightest chargino is nearly degenerate in mass with the lightest neutralino). Neutralino relic abundances for the three models considered, for a standard cosmological setup and no extra non-thermal sources, are shown in the left-hand side of Fig.~\ref{fig:oh2}, in the neutralino mass range allowed for each scenario.

In the right panel of Fig.~\ref{fig:oh2} we show instead an example of the quintessential enhancement of the thermal neutralino relic abundance at work on three models within the above described benchmark scenarios, at a common mass $m_\chi=300$ GeV. The upper panel shows the relic abundance of each model as a function of the parameter $\xi_\phi$, which is defined as the ratio between the quintessential energy density $\rho_\phi$ (most likely, in a "kination " phase, i.e. a phase of fast-rolling and $\rho_\phi$ scaling like $a^{-6}$, with $a$ the scale factor of the Universe) and the radiation energy density $\rho_{\rm rad}$ at the neutralino freeze-out temperature in the absence of quintessence. This parameter has been shown to suitably gauge out  most of the dependence on the parameters in the particle physics setup, and to describe the relic density enhancement effect in other cosmological scenarios, as for instance the case of an anisotropic primordial Universe with an effective shear energy density scaling again as $a^{-6}$~\cite{shear,Profumo:2004ex}. From the figure we deduce that, at a neutralino mass of 300 GeV, the {\bf Funnel} model would require an extra energy density component 10 times larger than that of radiation at neutralino freeze-out, while larger amounts are needed in the {\bf NUGM} and in the {\bf AMSB} cases (respectively around $10^3$ and $10^4$). The lower panel shows the relative enhancement, 
\begin{equation}
\Delta \Omega = \frac{\Omega_Q-\Omega}{\Omega}.
\end{equation}
As already mentioned and described in further details in ref.~\cite{Profumo:2003hq}, we find a rather small spread in values of $\Delta\Omega$ among the SUSY models considered here and for given values of the parameter $\xi_\phi$.

\section{Indirect searches through antimatter cosmic ray measurements} \label{sec:SUSYantimatter}

Since the first proposals to search for the exotic antimatter components in cosmic rays due to dark matter WIMP pair annihilations~\cite{silksre,SRW}, calculations of the expected fluxes have been performed with an increasing degree of sophistication. In recent estimates, see, e.g., refs.~\cite{bottino-pbar,pbarpaper,epluspaper,dbar, donato-pbar}, refinements have regarded both the modeling of source functions and the description of the propagation of charged cosmic rays in the Galaxy.

\subsection{Estimates of the antimatter fluxes}\label{sec:fluxestim}

We discuss here the case of neutralino-induced antiproton, positron and antideuteron cosmic ray fluxes. Predictions for the three species are derived in a consistent framework after going through a few steps. As already stressed, source functions are proportional to the annihilation probability for a pair of WIMPs, i.e. $\langle\sigma_{\rm eff}v\rangle(T=0)$, and depend on neutralino mass and on the branching ratios for the various annihilation channels. Annihilation yields fragment and/or decay, and generate stable antimatter species; for positrons and antiprotons, the simulation of this process is performed using the \code{Pythia} \cite{pythia} 6.154 Monte Carlo code, in each of the allowed two-body final state and for a set of 18 neutralino masses. For $\bar{D}$ sources we implement instead the prescription suggested in ref.~\cite{dbar} to convert from the $\bar{p}$-$\bar{n}$ yields. 

Source functions are proportional as well to the number of WIMP pairs which potentially may annihilate at each place in the Galaxy; postulating that dark matter is smoothly distributed in the Galactic halo, the number density of pairs can be expressed in terms of the dark matter density profile as $1/2\,(\rho_{\chi}(\vec{x}\,) / {m_{\chi}})^2$. Even in this limit, further extrapolations are needed, as, unfortunately, $\rho_{\chi}(\vec{x}\,)$ is poorly constrained by available dynamical data on the Galaxy. In this analysis, we will mainly focus on a dark matter halo described by the so-called Burkert profile~\cite{burkert}:
\begin{equation}
  \rho_{B}(r) = \frac{\rho_B^0}{(1+r/a)\,(1+(r/a)^2)}\;;
\label{eq:burk}
\end{equation}
it is a profile with a large core radius, that has been tested against a large sample of the rotation curves for spiral galaxies~\cite{salucci}. Such choice is very conservative, as results from N-body simulation of hierarchical structures in CDM cosmologies find instead singular halo profiles. In an extreme model for the baryon infall~\cite{elzant}, with very large angular momentum transfer between the baryonic and the dark components, a Burkert-type profile may replace the CDM cuspy halo after the gas has been settling in the inner portion of a galaxy to form its luminous components. Scenarios of baryon infall in which the CDM cusp is preserved are feasible as well, and we will discuss one such possibility in sec.~\ref{sec:hmodels}, mentioning also the implications for neutralino dark matter detection. We make a sample choice of the free parameters in eq.~\ref{eq:burk}, fixing the length scale parameter $a= 11.7$~kpc and the local halo density $\rho_B(r_0)= 0.34$~GeV~cm$^{-3}$; this configuration has been found after implementing available dynamical constraints, including constraints on the local halo density from the motion of stars in the Sun's neighborhood, on the total mass as derived from the orbit of the outer satellites, as well as by requiring consistency with the Milky Way rotation curve and with measures of the optical depth towards the galactic bulge. We have also taken into account the correlation between halo mass and halo concentration as found in the numerical simulations (in the structure formation language, the halo we consider has a virial mass and concentration parameter of, respectively, $M_{vir} = 1.3 \times 10^{12}\msun$ and $c_{vir} = 16$, see ref.~\cite{halomod} for details).

Once the source functions are fully specified, the next step is to model the propagation of charged cosmic rays through the Galactic magnetic fields. We consider an effective two-dimensional diffusion model in the steady state approximation. We do not include explicitly reacceleration effects in this propagation model, but mimic them through a diffusion coefficient which takes the form of a broken power law in rigidity, $R$, 
\begin{eqnarray}
D = D_0\,(R/R_0)^{0.6}\;\;\;\;\;\;\;\;& {\rm if}& R \ge R_0 \nonumber \\
D = D_0\;\;\;\;\;\;\;\;\;\;\;\;\;\;\; & {\rm if}& R < R_0\,.
\label{eq:diff}
\end{eqnarray}
This form has been used in a number of studies on the propagation of cosmic rays; e.g., in ref.~\cite{strmosk}, using the \code{Galprop}~\cite{galprop} propagation code, it has been shown that it gives a fair estimate of ratios of primary to secondary cosmic ray nuclei, for a suitable choice of the free parameters in the model. We implement here this same setup, i.e. take eq.~(\ref{eq:diff}) with $D_0 = 2.5 \times 10^{28}$~cm$^2$~s$^{-1}$ and $R_0=4$~GV, in a cylindrical diffusion region of radius equal to 30~kpc and half height equal to 4~kpc, plus a galactic wind term. For antiprotons and antideuterons, this setup is interfaced to the semi-analytic diffusive-convective propagation model described in ref.~\cite{pbarpaper}, which do not allow for an energy loss term (particles are removed whenever they scatter on the interstellar medium). For positrons, we exploit the
results in ref.~\cite{epluspaper}, where a propagation model with a term accounting for positron energy losses (the dominant terms accounting for inverse Compton scattering on starlight and the cosmic microwave background) was solved analytically; such model has been recently improved and extended to allow for the implementation of a diffusion coefficient in the form of eq.~(\ref{eq:diff}), and to keep a full two-dimensional structure
(the corresponding code is included in the latest release of the  \ds\ package). 

The final step to compare with measurements is to include a treatment of solar modulation effects, i.e. the propagation of cosmic rays through the solar system up to the Earth location, against the solar wind. To sketch this effect, we implement the 
one parameter model based on the analytical force-field approximation by Gleeson \& 
Axford \cite{GleesonAxford} for a spherically symmetric model. The solar modulation
parameter, sometimes dubbed Fisk parameter $\Phi_F$~\cite{fisk}, is for simplicity
assumed to be charge-sign independent, and it is assumed that, for all species, it can be chosen to be equal to the corresponding parameter as determined for proton cosmic-ray flux. The latter has been measured with some accuracy over the full 11-year solar cycle.

\FIGURE[t]{\epsfig{file=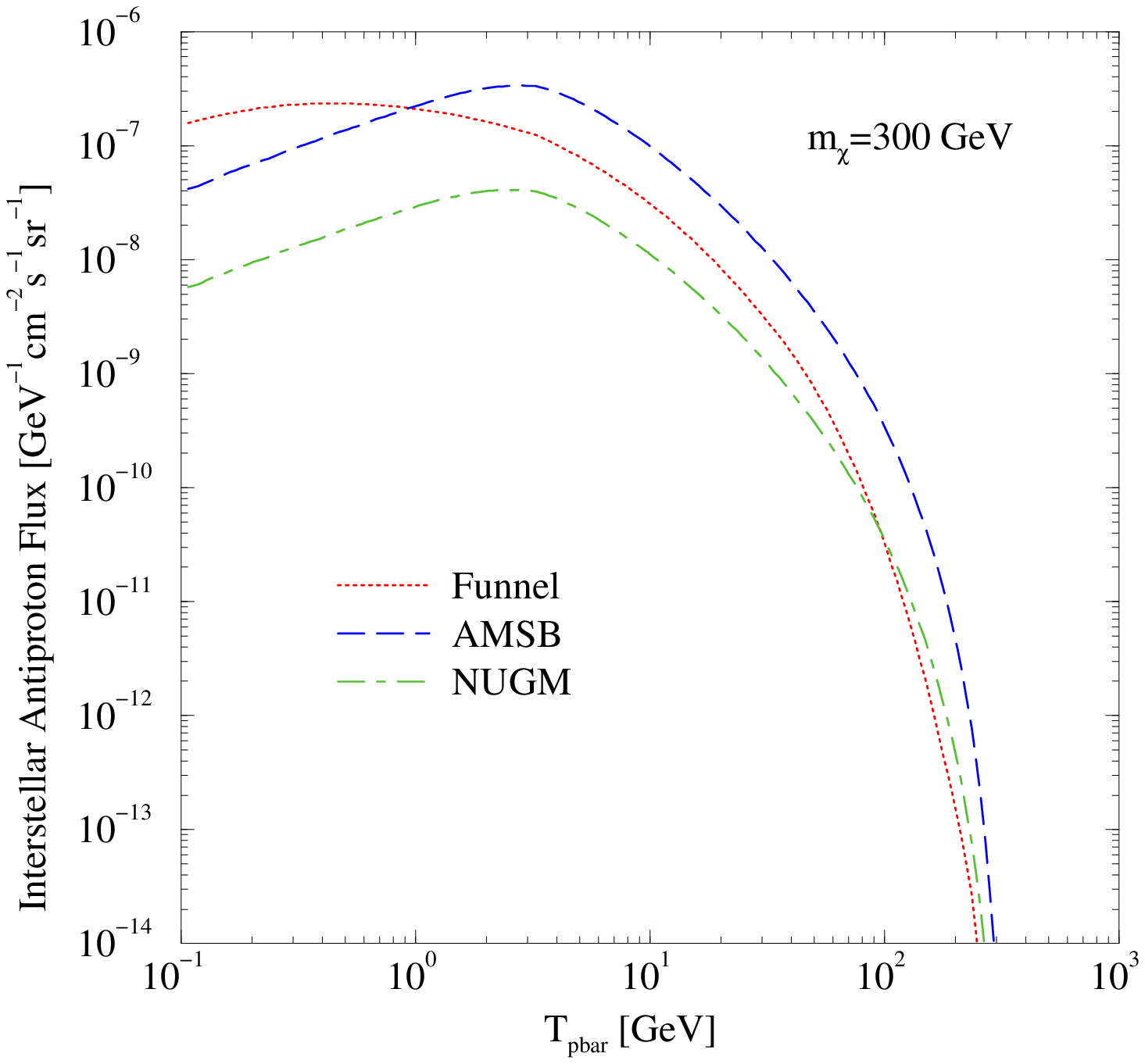,width=7.1cm}\qquad
\epsfig{file=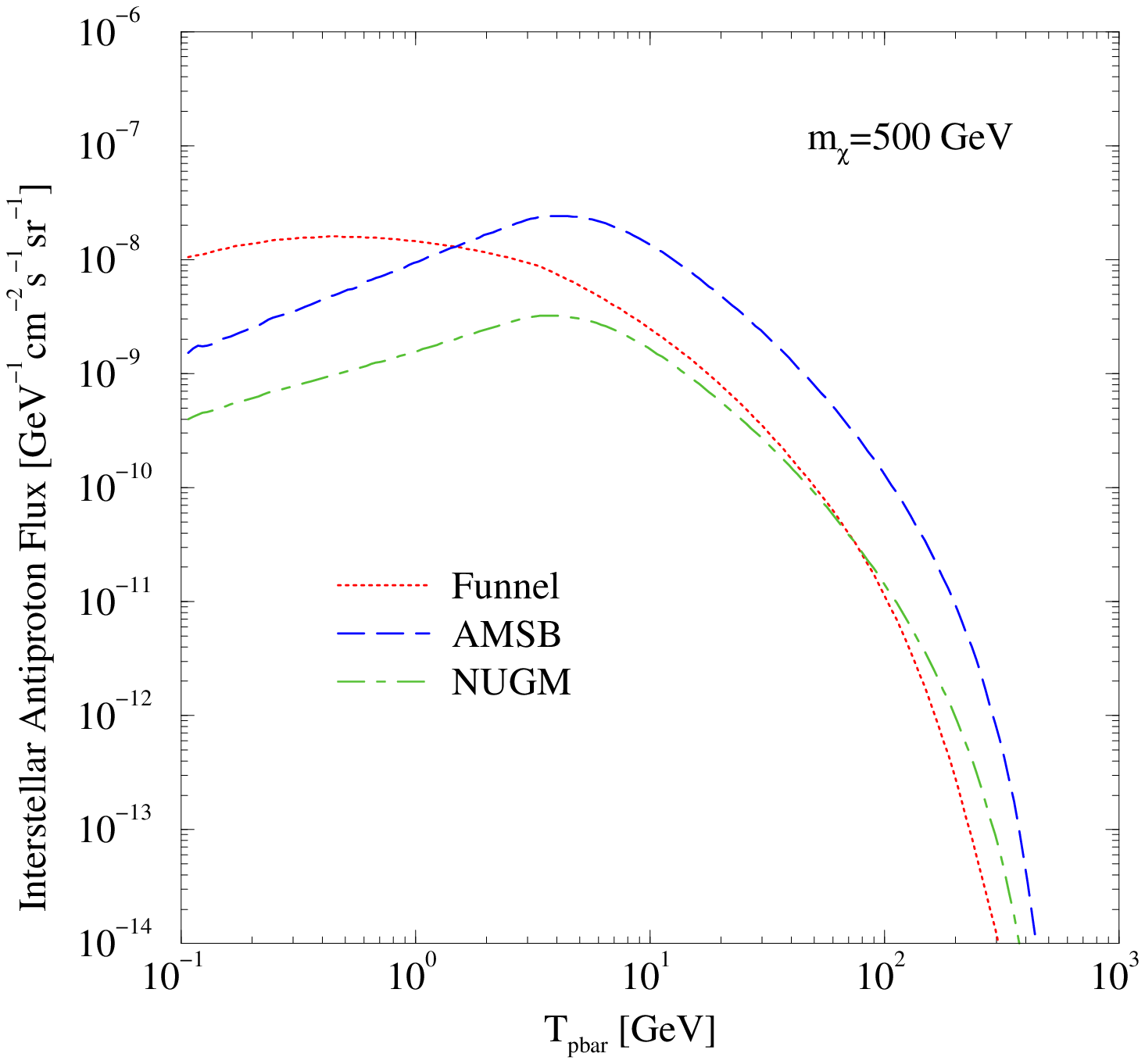,width=7.1cm} 
\caption{The interstellar antiprotons flux, as a function of the antiprotons kinetic energy $T_{\rm pbar}$, for two values of the neutralino mass, respectively $m_\chi=300$ GeV (left) and $m_\chi=500$ GeV (right), for the three SUSY models of Tab.~\ref{tab:models}.}
\label{fig:pbIS}}

\subsection{Antimatter spectral features for the three benchmark models}

Antimatter fluxes from neutralino annihilations depend on one hand on the number density of neutralinos in the galactic halo (and therefore, for a given CDM halo profile, on the inverse of the neutralino mass squared), and on the other hand  on the specific features of the annihilation processes of neutralinos, in terms of the overall interaction rate and of the standard model final products. The three benchmark models considered in the present work feature relatively large annihilation cross sections, when compared to the rule of thumb estimate, approximately valid in a standard cosmological scenario, 
\begin{equation}
\langle \sigma_{\rm eff} v\rangle\simeq \frac{3\cdot 10^{-27} \,{\rm cm}^3\,{\rm s}^{-1}}{\Omega_{\rm CDM}h^2} 
\end{equation}
 The values of $\langle \sigma_{\rm eff} v\rangle$, at a sample neutralino mass of 300 GeV, lie in fact between $10^{-24}$ and $10^{-25}$ ${\rm cm}^3{\rm s}^{-1}$. As expected, low relic density models will thus have noticeably large antimatter fluxes, as compared to models which give the expected relic abundance within the standard cosmological scenario.

\FIGURE[t]{\epsfig{file=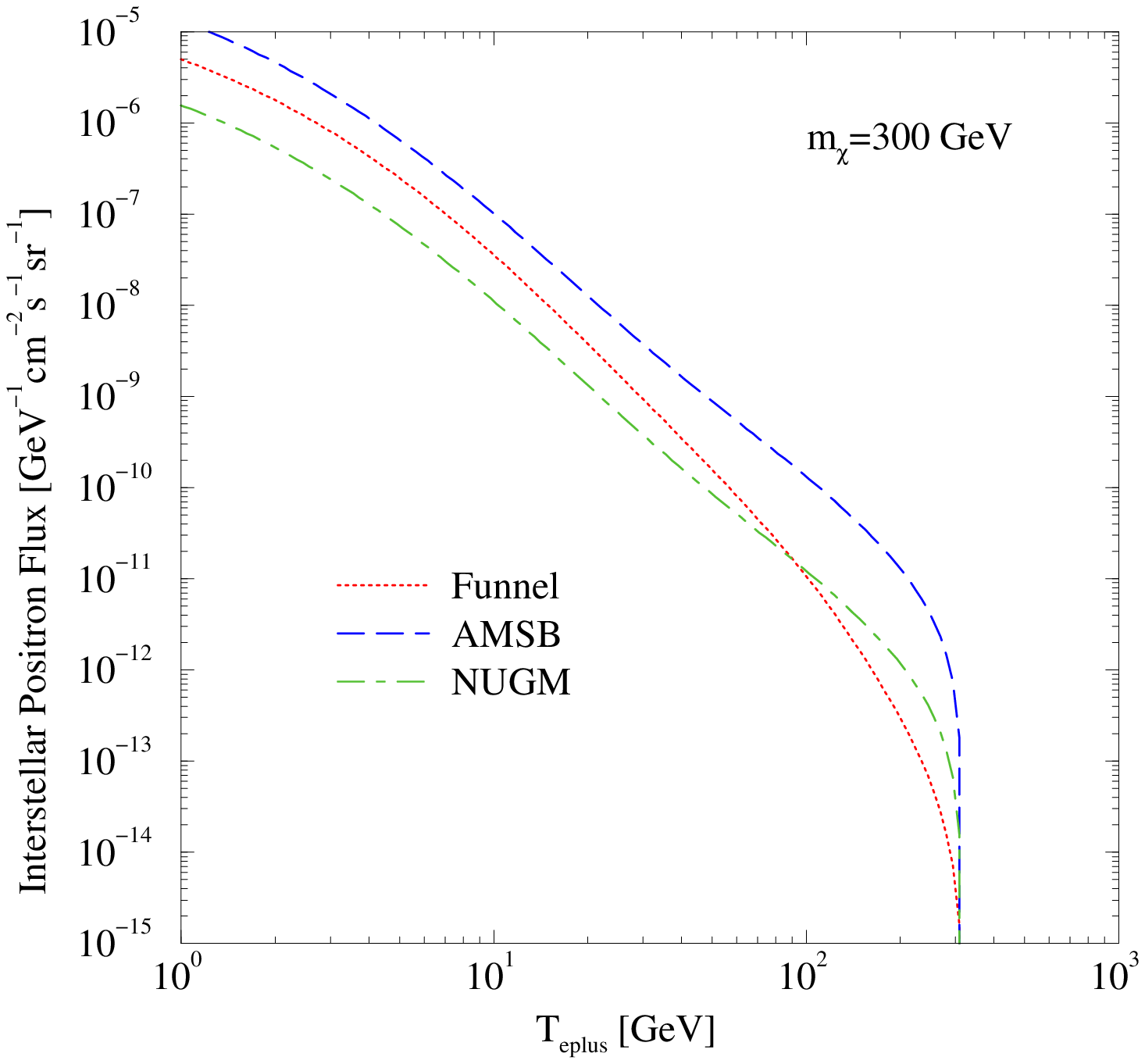,width=7.1cm}\qquad
\epsfig{file=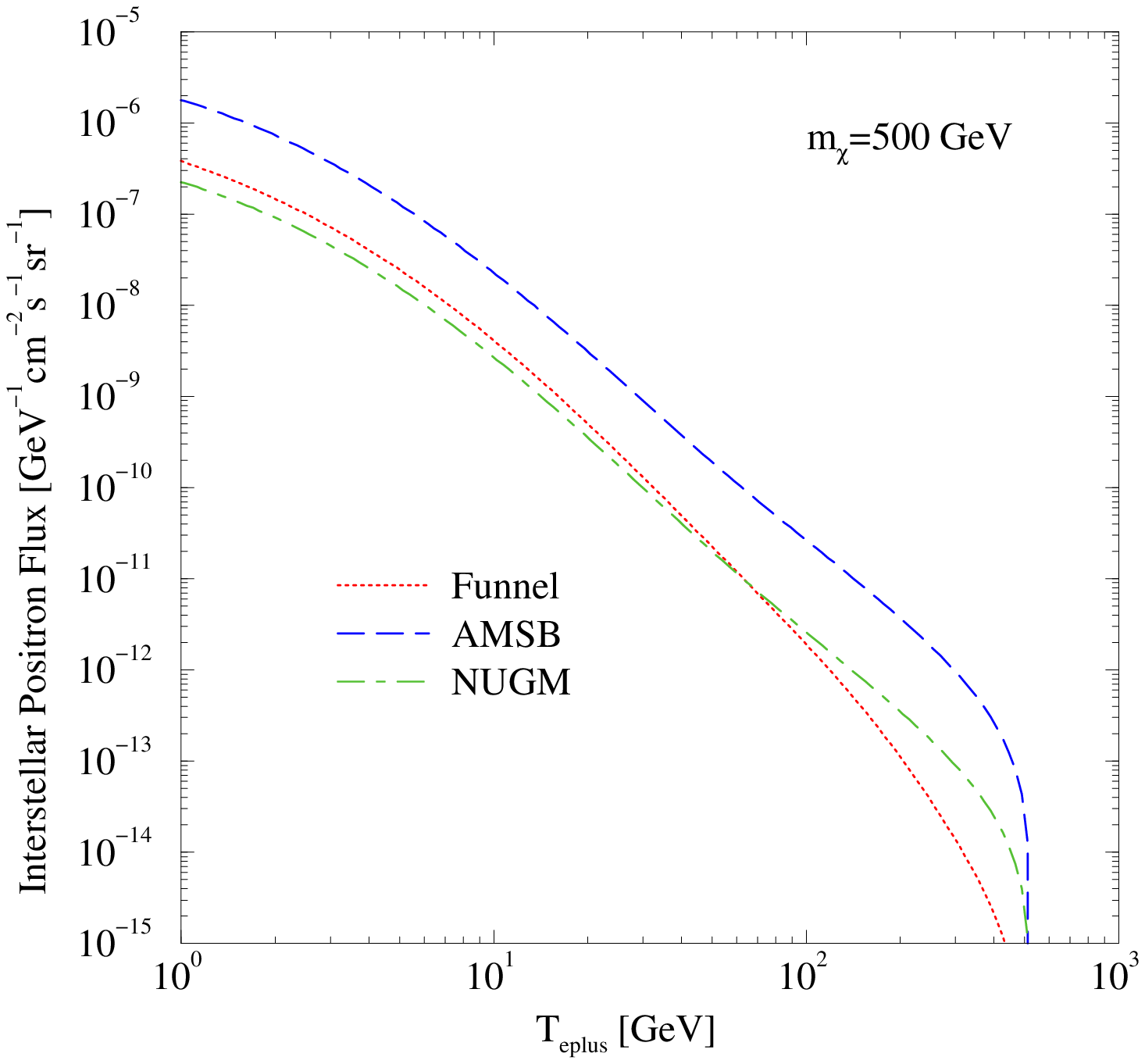,width=7.1cm} 
\caption{The interstellar positrons flux, as a function of the positrons kinetic energy $T_{\rm eplus}$, for two values of the neutralino mass, respectively $m_\chi=300$ GeV (left) and $m_\chi=500$ GeV (right), for the three SUSY models of Tab.~\ref{tab:models}.}
\label{fig:epIS}}

We show in figures \ref{fig:pbIS}, \ref{fig:epIS} and \ref{fig:dbIS} the interstellar fluxes of antiprotons, positrons and antideuterons for a few selected models within the three benchmark scenarios. In the left and right panels of each of the figures we compare (kinetic) energy spectra for the given antimatter species at different neutralino masses, respectively 300 GeV (left) and 500 GeV (right). The $1/m_\chi^2$ suppression effect due to neutralino number density dilution at growing masses is everywhere apparent.

The spectral features of the antimatter fluxes, on the other hand, critically depend on the decay channels of the annihilating neutralinos. In the case of the {\bf Funnel} model, the largely dominating decay channels are $b\overline{b}$ (around 90\%) and $\tau^+\tau^-$ (around 10\%): this pattern is due to the fact that the value of $\tan\beta$ which has been chosen is particularly large, so that decays into up-type fermions (charm or top) are very suppressed with respect to those into down-types. The resulting antimatter from hadronization of $b$ quarks as well as $\tau$ decay products yield typically {\em soft} antimatter particles: as emerging from the figures, at the lowest energies of interest, binos in the {\bf Funnel} model always give larger fluxes with respect to higgsinos and winos.

\FIGURE[t]{\epsfig{file=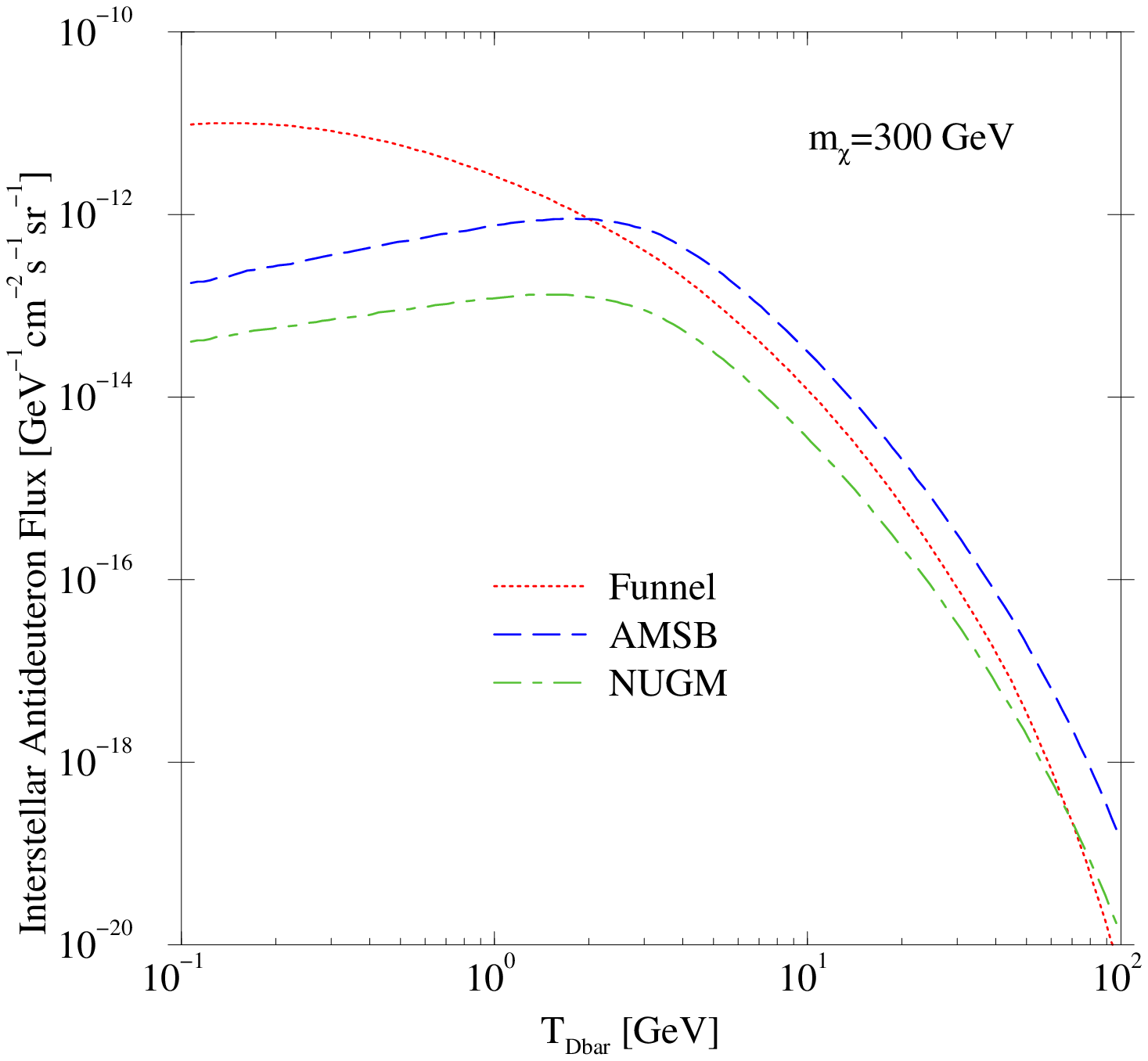,width=7.1cm}\qquad
\epsfig{file=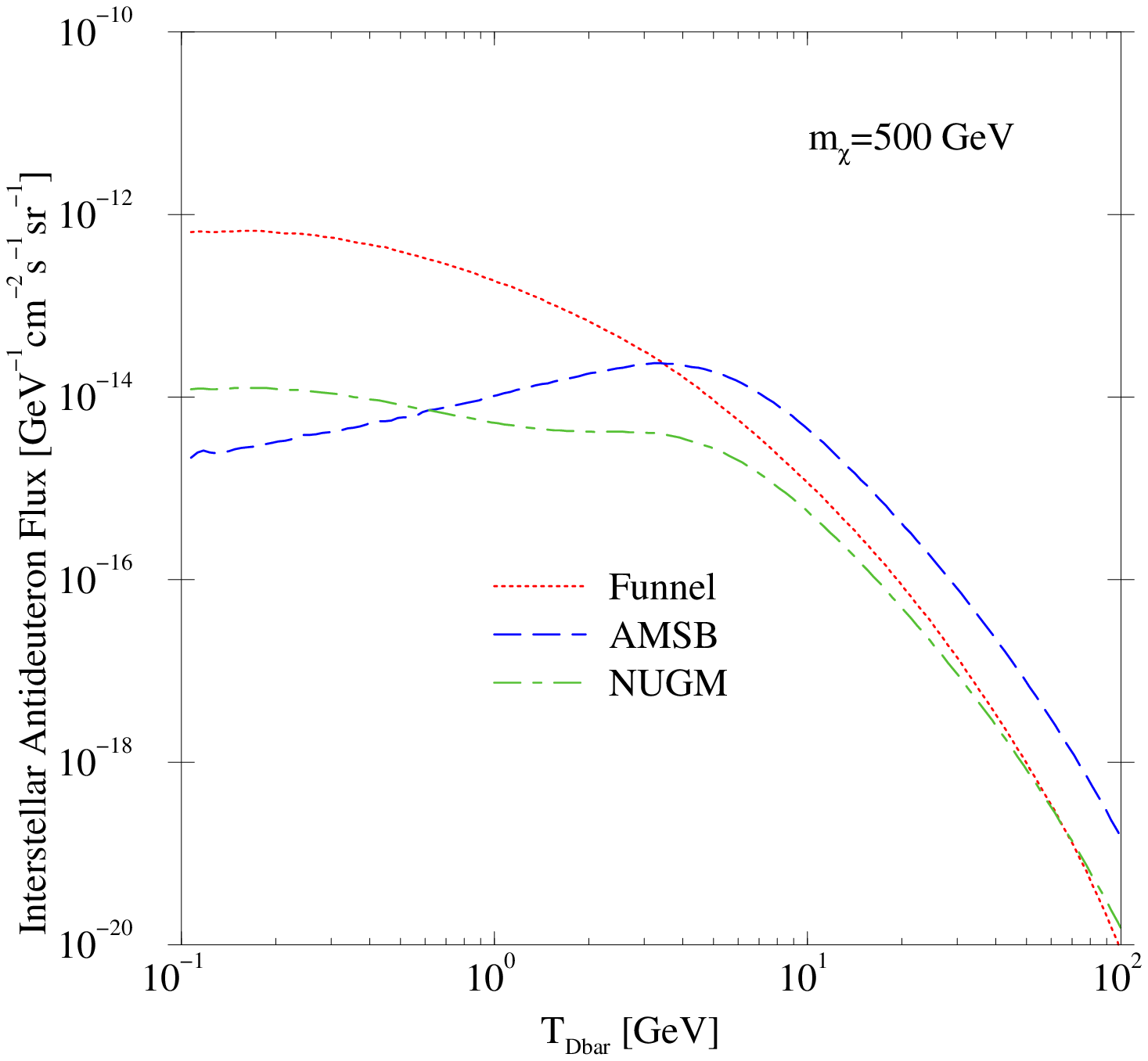,width=7.1cm} 
\caption{The interstellar antideuterons flux, as a function of the antideuterons kinetic energy $T_{\rm Dbar}$, for two values of the neutralino mass, respectively $m_\chi=300$ GeV (left) and $m_\chi=500$ GeV (right), for the three SUSY models of Tab.~\ref{tab:models}.}
\label{fig:dbIS}}

The energy spectrum in the {\bf AMSB} model reflects the large degree of purity of the dominating decay channel, namely that in $W^+W^-$, whose branching ratio is always above 98\%. As regards hadronic antimatter (antiprotons and antideuterons), annihilating winos tend to give rise, from hadronization of $W$ decay products, to a peak in the antimatter particle energy around a few GeV. The precise location of the peak non-trivially depends on the details of the antiproton and antineutron production, though its location in energy {\em increases} at larger masses (see fig.~\ref{fig:pbIS} and \ref{fig:dbIS}). In the positron channel, a clean peak is located at $E_{e^+}\simeq m_\chi/2$: the origin of this very neat feature is the prompt decay $W^+\rightarrow e^+\ \nu_e$, where, since the annihilating neutralinos are non-relativistic, the energy of the decaying $W$ is close to $m_\chi$.

\FIGURE[t]{\epsfig{file=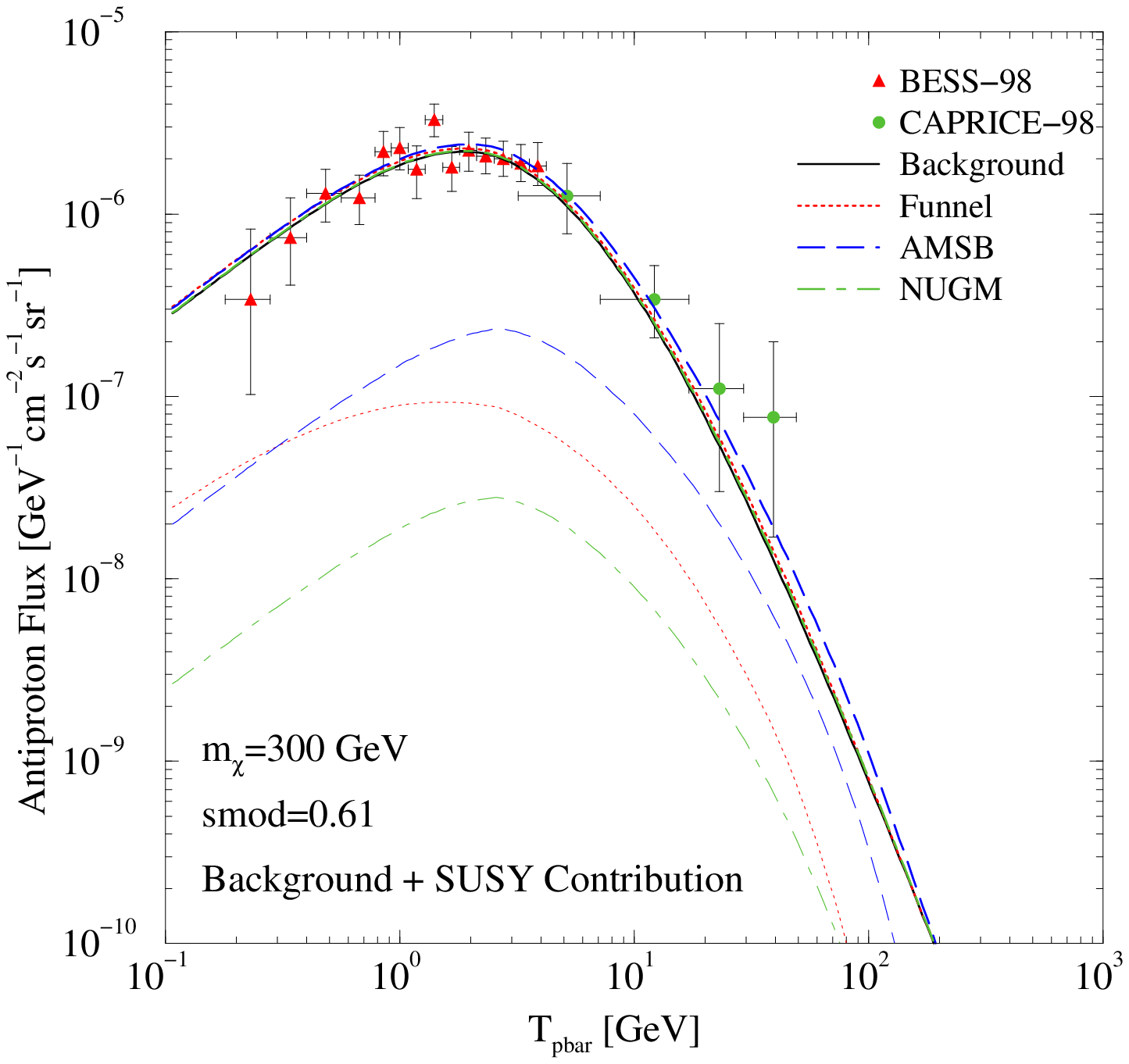,width=7.1cm}\qquad
\epsfig{file=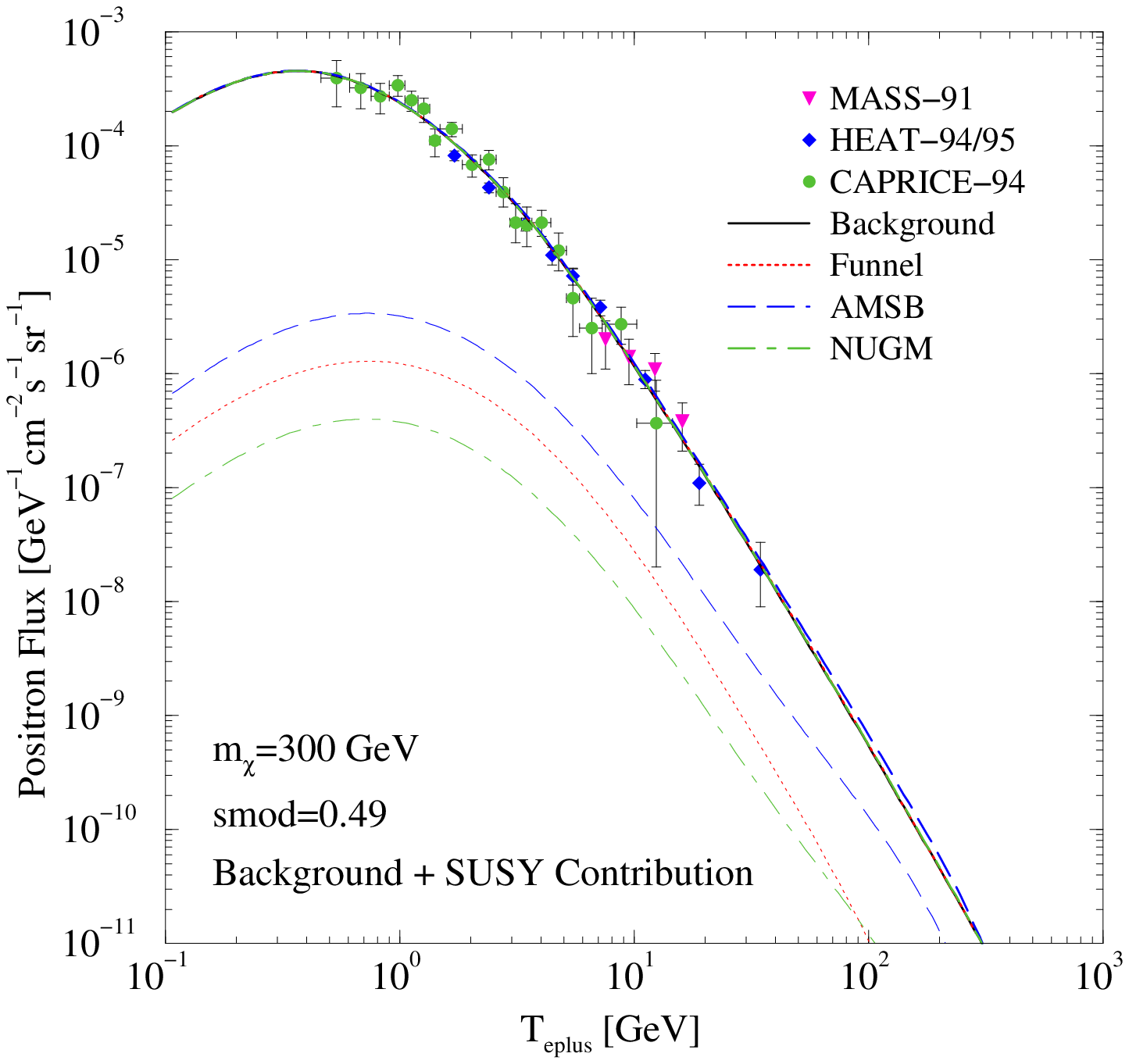,width=7.1cm}
\caption{(a): The solar modulated antiprotons flux, as a function of the antiprotons kinetic energy $T_{\rm pbar}$. The black line corresponds to the calculated background, while the three colored thick lines to the total signal for the three SUSY models at a mass  $m_\chi=300$ GeV. The thin lines correspond to the SUSY contributions alone. The data from BESS-98 \cite{bess} and CAPRICE-98 \cite{capricepbar} are also shown. (b) The solar modulated positrons flux, as a function of the positrons kinetic energy $T_{\rm eplus}$. The black line corresponds to the calculated background, while the three colored thick lines to the total signal for the three SUSY models at a mass  $m_\chi=300$ GeV. The thin lines correspond to the SUSY contributions alone. We also plot the positron data from MASS-91 \cite{mass91}, HEAT-94/95 \cite{heat} and CAPRICE-98 \cite{capriceeplus}.}
\label{fig:pbSM}}

Turning to the case of higgsinos ({\bf NUGM} model), the spectral features of antiprotons and positrons neatly reproduce those of winos, though at lower fluxes (the suppression being motivated both by a smaller effective annihilation cross section and by a suppression factor in the couplings of higgsinos to the $W$ bosons). Noticeably, the antideuteron flux is instead larger than that of winos at energies below the GeV. This pattern is motivated by the fact that higgsinos decay with $\sim50$\% probability into $W^+W^-$, and with $\sim45$\% probability into $Z^0Z^0$: while antiprotons and positrons production from $Z^0$ decays is very similar to that from $W$ decays, the creation of low energy antineutrons from $Z^0Z^0$ is remarkably enhanced, thus giving rise to the mentioned peak in the antideuteron low-energy flux.

Fig.~\ref{fig:pbSM} shows the spectral features, after solar modulation for a given step along the solar activity cycle (see the discussion in sec.~\ref{sec:fluxestim}), of primary antiparticles and of the background, comparing the total expected signals to the data on antiprotons and positrons taken during the corresponding modulation phase. The figure refers to a common neutralino mass of 300 GeV. As regards the secondary antimatter 
fluxes, which play here the role of backgrounds, our estimates are produced running
the  \code{Galprop}~\cite{galprop} code in the configuration for propagation parameters
we have adopted for the signals. We remark that for both species, the computed backgrounds provide by themselves excellent fits of the data: we obtain,
for background only, a reduced $\chi^2$ equal to 0.82 for antiprotons and to 0.95
for positrons. A primary flux will then be {\em excluded} as long as it gives rise to a statistically unacceptable $\chi^2$, no longer compatible with currently available data. At a given neutralino mass, the signal-to-background ratio for antiprotons, after solar modulation, is much more promising than that of positrons, with the possible exception of large energies, around the $m_\chi/2$ peak in the positron spectrum, which lies however far above the currently sampled energies.

\section{Current exclusion limits}\label{sec:curlim}

Ruling out a model from antimatter fluxes amounts to evaluate the $\chi^2$ of the expected signal plus background, after taking care of the different solar modulation effects in the relevant period of data-taking. We will show here the {\em reduced} $\chi^2$, i.e. the $\chi^2$ divided by the relevant number of data (respectively 49 for antiprotons and 32 for positrons). The 95\% confidence level (C.L.) exclusion limit will lie at a $\chi^2$ around 1.33 for antiprotons and 1.4 for positrons, though $\chi^2\sim 1$ indicates that the model is around its ``{\em visibility threshold}''. We will find models for which reduced $\chi^2$ are above 10 or so, which are excluded with no doubt, and models closer to the quoted confidence levels; for the latter one should keep in mind that we are not taking into account uncertainties in the propagation parameter model and in the halo profile (although in this respect we are taking a rather conservative scenario), hence limits we will quote should not be intended as sharp cut-offs. 

The quality of the data on the local antiproton and positron cosmic ray flux has kept
improving in recent years. As regards the antimatter flux data we take into account,
we will compare the predicted fluxes to the antiproton data 
collected by the BESS experiment during its flights in 1997, 1998, 1999 and 2000
\cite{bess}, with fairly good statistics, in the energy range between 180 MeV and 
4.2 GeV, and by the CAPRICE experiment during its 1998 flight~\cite{capricepbar} 
in the range between 3 and 50 GeV\@. For the positron fluxes, we consider the data published
by the  HEAT Collaboration about the 1994-1995 flight~\cite{heat}, by the CAPRICE team 
in a flight in 1994~\cite{capriceeplus}, and finally data obtained by MASS-91 in 
1991 \cite{mass91}: the overall energy range covered by these measurements extends from
460 MeV to 34.5 GeV. We have chosen not to include in our analysis data which
have been reported just as antiproton or positron fractions (rather than absolute 
fluxes) and datasets such as the one on positrons from the AMS test 
flight \cite{amseplus} mapping a low energy interval in which a primary 
neutralino-induced contribution is expected to be rather suppressed. 

\FIGURE[t]{\epsfig{file=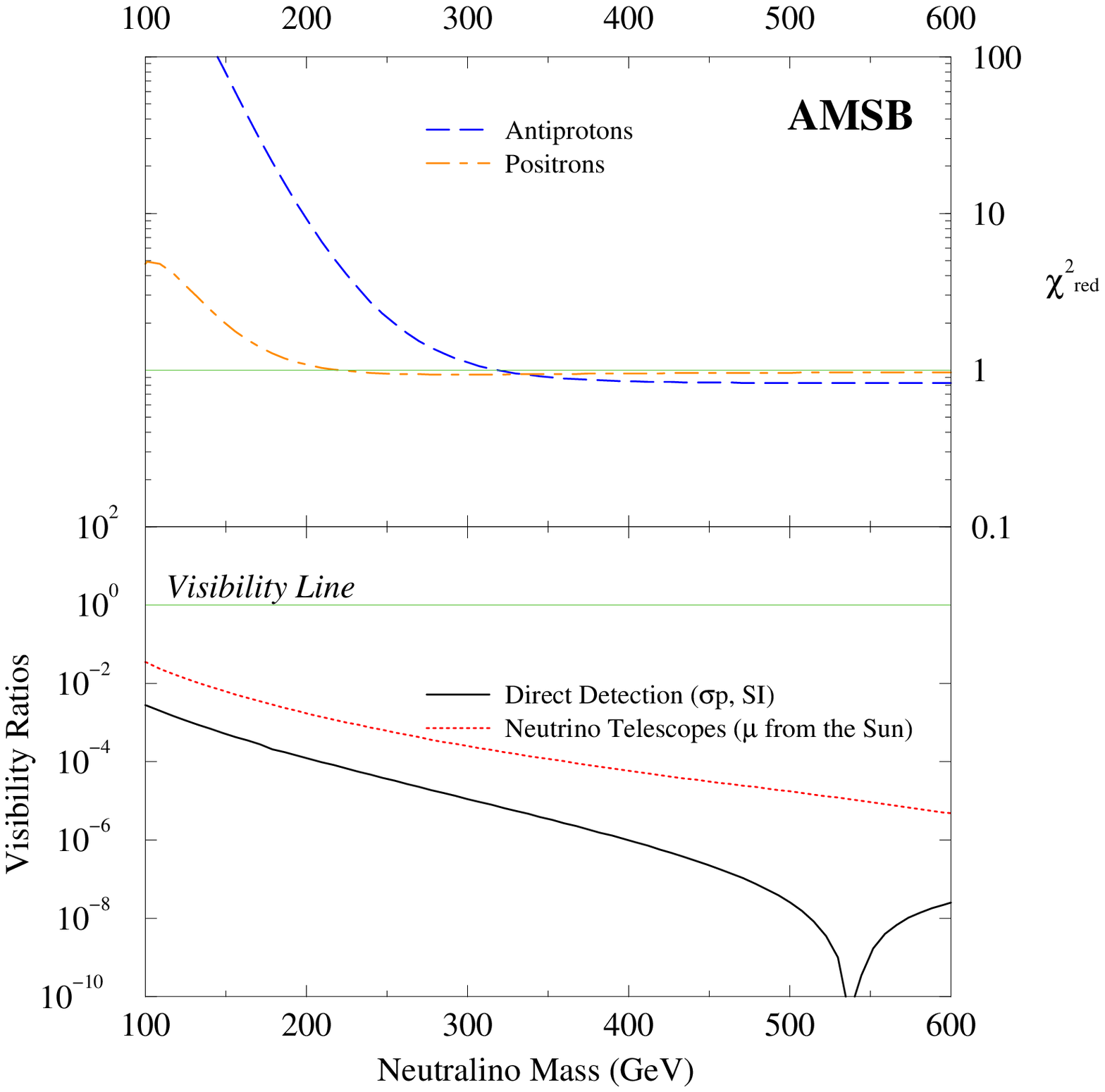,width=10.5cm}
\caption{Current exclusion limits on the {\bf AMSB} model from direct and indirect Dark Matter searches, as a function of the neutralino mass. In the upper panel we show the reduced $\chi^2$ of the background plus SUSY signal, and, as a guideline, $\chi^2_{\rm red}=1$; in the lower panel we show the ratio of the expected signal and of the current exclusion limits ({\em Visibility Ratio}) for direct dark matter detection (neutralino-proton spin-independent scattering cross-section, solid line) and on the neutrino-induced muon flux from neutralino pair-annihilation in the center of the Sun (dotted line). The ``{\em Visibility Line}'' corresponds to values of the expected signal equal to the current exclusion limits: models above the Visibility Line would be already ruled out.}
\label{fig:CurrentAMSB}}

\FIGURE[t]{\epsfig{file=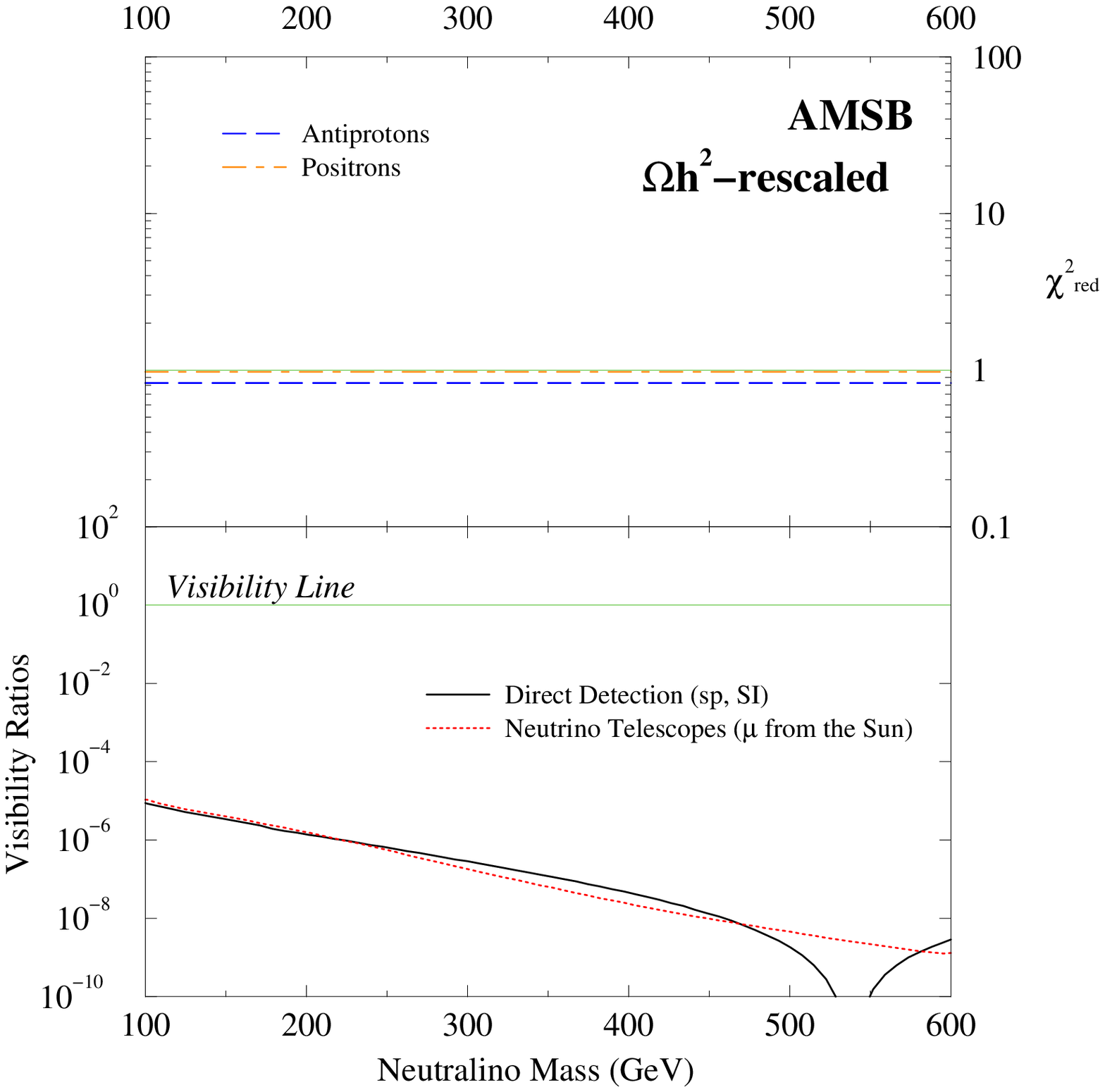,width=10.5cm} 
\caption{The same as in fig.~\ref{fig:CurrentAMSB}, again in the {\bf AMSB} model, but with a rescaled neutralino density profile, according to the formula \mbox{$\rho_\chi^{\rm resc}=\rho_{\rm\sss CDM}\ \Omega_{\rm\sss CDM}/{\rm Min}\left(\Omega_\chi,\Omega^{\rm min}_{\rm\sss CDM}\right)$}.}
\label{fig:CurrentAMSB_resc}}

We compare the statistical analysis on antimatter fluxes with the expected signal-to-sensitivity (``{\em visibility}'') ratio for direct, spin-independent searches and for the muon flux at neutrino telescopes originating from neutralino annihilations in the center of the Sun\footnote{We always find that spin-dependent direct detection, and the muon flux from the center of the Earth or from the Galactic center have visibility ratios by far smaller than those we consider.}. The visibility ratio for direct detection refers to the expected neutralino-proton scattering cross section over the current experimental sensitivity level. 
For definiteness, we will refer to the EDELWEISS  2002~\cite{edelweiss} results and estimate the corresponding exclusion limit, rather than within the standard Maxwell-Boltzmann local velocity distribution as usually done, by using the {\em self-consistently} derived velocity distribution of the halo model under consideration, and taking into account relevant effects, such as target materials, form 
factors, and threshold.~\footnote{An effect compatible with being due to WIMP-nucleon
scatterings has been reported by the DAMA Collaboration, see their latest report in 
ref.~\cite{dama}. Unfortunately, none of the models considered here give an effect at the level of such signal.}

As far as the muon flux induced by neutrinos from the sun, the current best limits are from the SUPER-KAMIOKANDE Collaboration in 2002~\cite{superKlimit}. Also for this detection method, the signal is computed estimating capture rates with the appropriate velocity distribution.

\subsection{The {\bf AMSB} model}

\FIGURE[t]{\epsfig{file=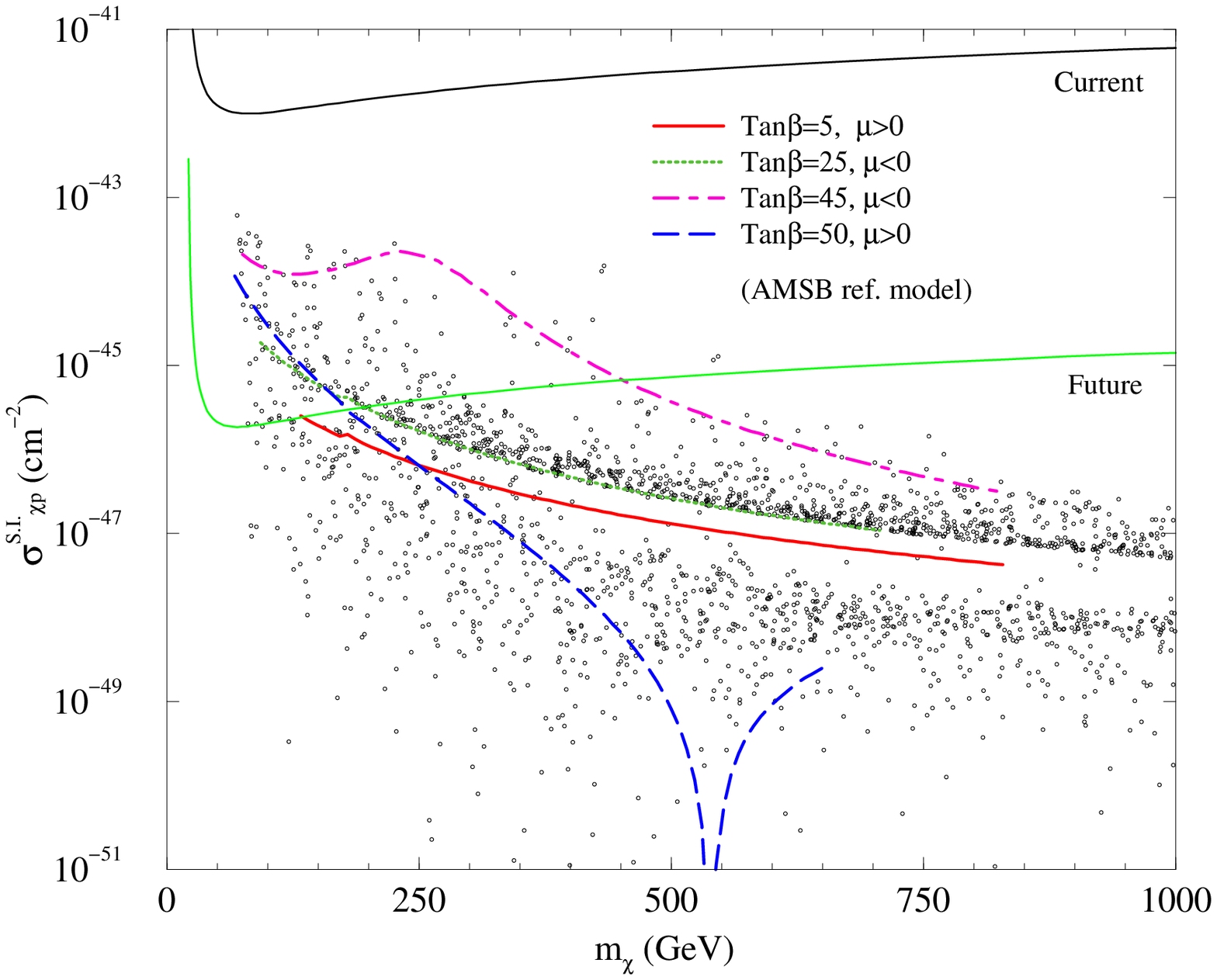,width=7.1cm}\qquad
\epsfig{file=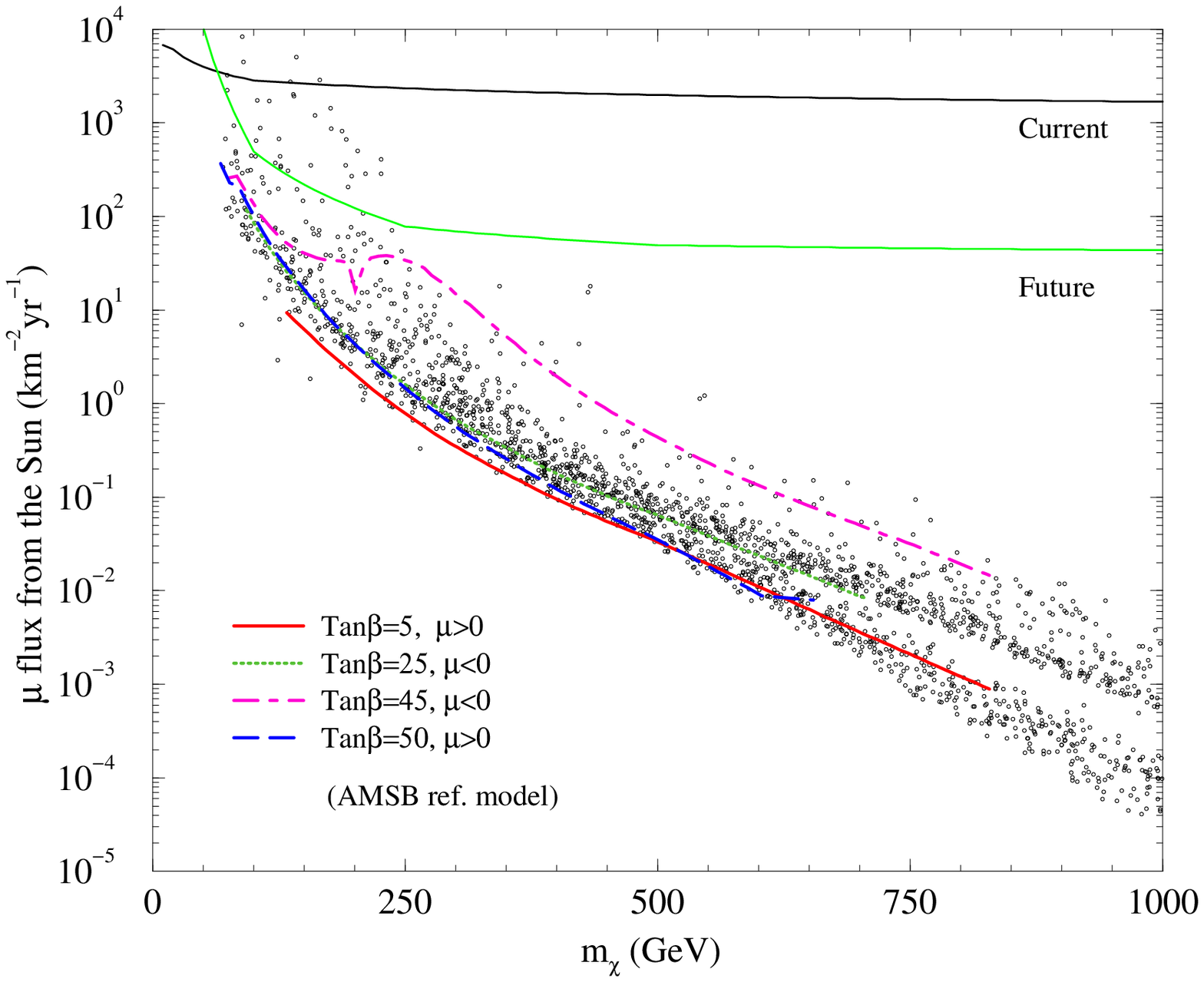,width=7.1cm} 
\caption{Direct spin independent neutralino-proton scattering cross section (left) and Muon flux induced by neutrinos produced by neutralino annihilations in the center of the Sun (right) for various (around 2500) minimal Anomaly Mediated SUSY breaking models. We linearly scanned $2<\tan\beta<50$, $200<m_0<5000$ and $10^4<m_{3/2}<4\cdot 10^5$ and picked with equal probability either sign of $\mu$. Four particular parameter space slices at $m_0=1500$ GeV, and at different $\tan\beta$ and sign of $\mu$ are also shown as guidelines.}
\label{fig:AMSBddid}}

Fig.~\ref{fig:CurrentAMSB} shows the current visibility ratios and exclusion limits for the {\bf AMSB} model. Remarkably, we find that direct detection, in this particular case, is the less promising search strategy, with expected signals lying more than two orders of magnitudes below current sensitivity. Neutrino telescopes do little better. On the other hand, masses below $\sim270$ GeV are {\em ruled out}, within the conservative Burkert halo model we consider, from the overall fit to antiprotons data, while the expected positron flux is above current sensitivity for masses below approximately 170 GeV. In case we had assumed that the model under scrutiny is describing some {\em subdominant} component of the full CDM term, and if we had accordingly rescaled the neutralino densities according to the thermal relic abundance computed in a standard cosmological scenario, i.e. applying the rescaling recipe:
\begin{equation}
 \rho_\chi^{\rm resc}=\rho_{\rm\sss CDM}\ \frac{\Omega_{\rm\sss CDM}}{{\rm Min}\left(\Omega_\chi,\Omega^{\rm min}_{\rm\sss CDM}\right)}\;,
\end{equation}
results would be largely altered, as we show in fig.~\ref{fig:CurrentAMSB_resc}: direct detection and neutrino telescope visibility ratios would have been both five orders of magnitude below one, and antiprotons and positron fluxes would give rise to a signal completely indistinguishible from the expected background. Relic density enhancement mechanisms are therefore {\em mandatory} within the minimal AMSB framework to achieve signals compatible with current, as well as with future experimental sensitivity. We find that the same conclusion applies for the other benchmark models we consider here.

As manifest from the shape of the corresponding curve in fig.~\ref{fig:CurrentAMSB}, direct detection rates are particularly low because a cancellation in the contributions from the $t$-channel exchanges of $H_1^0$ and $H_2^0$ takes place in the particular parameter slice we consider here as a {\em benchmark} scenario. For clarity, we performed a random scan of the full mAMSB parameter space (see the caption in fig.~\ref{fig:AMSBddid} for details on the scan). We report in the scatter plots of fig.~\ref{fig:AMSBddid} our results, together with three parameter space slices as guidelines and with the considered {\bf AMSB} model as well. The left part of the figure refers to direct detection, while on the right we display muon fluxes at neutrino telescopes. Notice that, although larger scattering cross sections may be obtained, all models are far below current sensitivity for direct detection, therefore not altering our conclusions. As regards neutrino telescopes, though very few points lie above the current exclusion limits, we point out the the resulting masses are always smaller than the limits we derived from current data on antimatter searches.

\subsection{The {\bf Funnel} and the {\bf NUGM} models}

Current exclusion limits for the {\bf Funnel} and for the {\bf NUGM} models are showed in figures \ref{fig:CurrentFUNN} and \ref{fig:CurrentNUGM} respectively. In the case of the {\bf Funnel}, where masses are rather large, we do not find any point which is neither ruled out by current data on antimatter fluxes nor by direct detection and neutrino telescopes. The latter two experimental techniques have achieved so far a sensitivity three orders of magnitude or worse below the expected signals. 

\FIGURE[t]{\epsfig{file=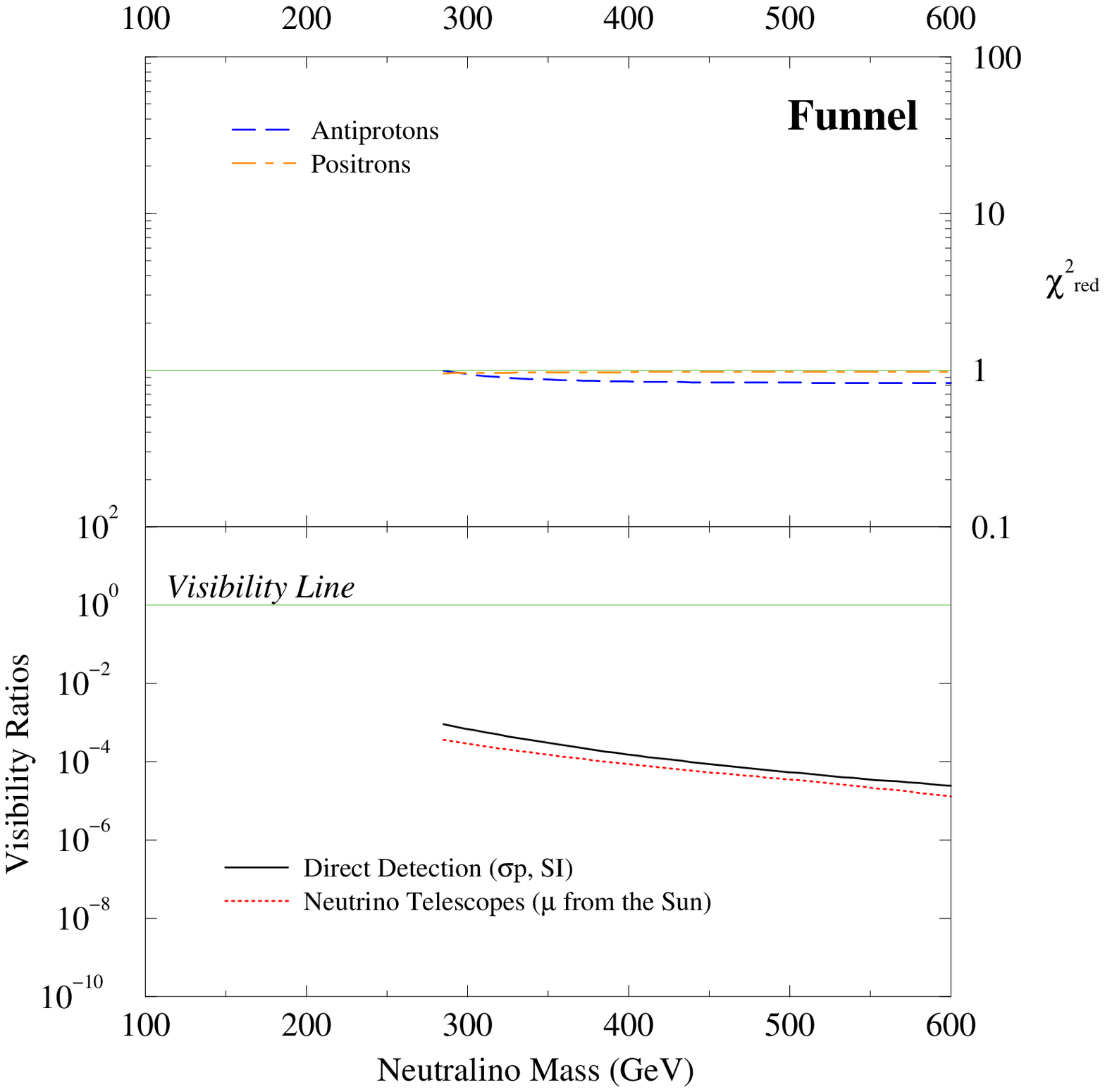,width=10.5cm}
\caption{Current exclusions limits on the {\bf Funnel} model. Symbols and conventions are the same as in fig.~\ref{fig:CurrentAMSB}.}
\label{fig:CurrentFUNN}}

\FIGURE[t]{\epsfig{file=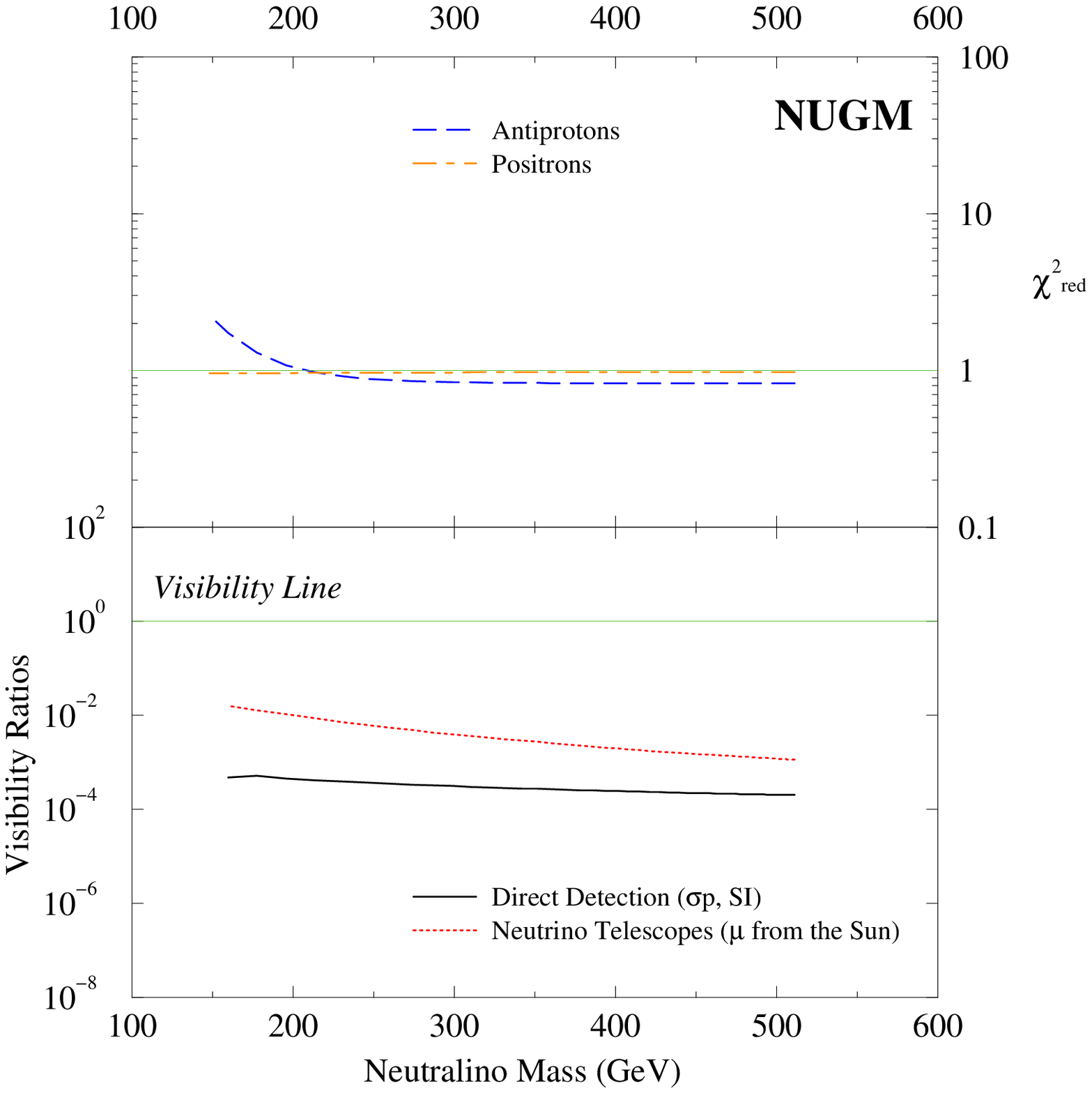,width=10.5cm} 
\caption{Current exclusions limits on the {\bf NUGM} model. Symbols and conventions are the same as in fig.~\ref{fig:CurrentAMSB}.}
\label{fig:CurrentNUGM}}

Regarding the {\bf NUGM} model, once again we single out a case for which direct dark matter detection is the least competitive detection strategy. Neutrino telescopes, though featuring a one order of magnitude better visibility ratios, are also a factor 100 or more below the needed sensitivity. Turning to antimatter, in the {\bf NUGM} scenario positrons do not give any statistical constraint, while antiproton expected fluxes are not compatible with current data for low masses. We remark that this is another example of a SUSY model which is currently constrained {\em only} by antimatter searches; moreover notice that this is true with the rather conservative Burkert halo profile considered here, and that constraints from antimatter searches, within other profiles, can be much tighter (see sec.~\ref{sec:hmodels}).

\section{Prospects for SUSY DM detection through antimatter searches}\label{sec:futlim}

\subsection{Discrimination perspectives at future antimatter search experiments}

New generation space-based experiments for antimatter searches PAMELA~\cite{pamela} and AMS~\cite{ams} will tremendously enhance the resolution and accuracy of positron and antiproton spectra measurements, as compared to existing balloon borne results (see sec.~\ref{sec:SUSYantimatter}). With the purpose of assessing discrimination capabilities of future experimental facilities, we will sketch here the possibility of disentangling an exotic component out of a standard secondary background. To this extent, we will implement a statistical $\chi^2$ analysis to compare the case of a pure background measurement to that of the occurrence of a signal. 

The relevant experimental parameters entering the estimate are given by:
\begin{itemize}
\item[-] The {\em geometrical factor} of the experimental facility, i.e. its effective area, $A$; 
\item[-] The time of data acquisition $T$;
\item[-] The energy coverage of experiment, with the relative definition of energy bins, i.e. its number $n_{\rm b}$ and the size $(\Delta E)_i$ of each of them.
\end{itemize}

We will declare that a given SUSY model is going to be {\em discriminable} at a certain future experiment, and at given X\% confidence level, if the $\chi^2$ induced by the SUSY model is larger than the $(\chi^2)^{X\%}_{n_{\rm b}}$ corresponding to $n_{\rm b}$ degrees of freedom. Letting $N_i^P=N_i^S+N_i^B$ be the number of projected events in a given bin $i$, the sum of the number of signal plus the background events, and $N_i^O$ be the number of observed events, with a standard deviation $\Delta_{N_i^O}$, the $\chi^2$ is defined as 
\begin{equation}\label{eq:chin}
\chi^2=\sum_{i=1}^{n_{\rm b}}\frac{\left(N_i^P-N_i^O\right)^2}{\left(\Delta_{N_i^O}\right)^2},\, \, \, \, N_i^P=N_i^S+N_i^B. 
\end{equation}
We will suppose that the standard deviation has a Gaussian distribution, i.e. 
\begin{equation}
\Delta_{N_i^O}\simeq \sqrt{N_i^O}.
\end{equation}
We are interested in finding the limiting cases, i.e. those cases for which an eventual signal is a small component with respect to the background. We will therefore make the assumption that:
\begin{equation}\label{eq:approximation}
N_i^S\ \ll N_i^B\;,\;\;\;\;{\rm or}\;\;\;\;N_i^O\ \simeq\ N_i^B\;.
\end{equation}
Eq.~(\ref{eq:chin}) will then read
\begin{equation}\label{eq:chinbis}
\chi^2=\sum_{i=1}^{n_{\rm b}}\frac{\left(N_i^S\right)^2}{\left(\sqrt{N_i^B}\right)^2}
\end{equation}
Now, since the number of events in an energy bin $\Delta E$ is given, as a function of the flux of particles $\phi$, by
\begin{equation}
N=(\Delta E)\cdot\phi\cdot A\cdot T,
\end{equation}
and indicating with $\phi_s$ and $\phi_b$ the signal and background fluxes respectively, eq.~(\ref{eq:chinbis}) will read
\begin{equation}\label{eq:chisquareddef}
\chi^2=\sum_{i=1}^{n_{\rm b}}\ \frac{\left(\phi_s^i\right)^2}{\phi_b}\cdot(\Delta E)_i\cdot A\cdot T.
\end{equation}
The quantity in eq.~\ref{eq:chisquareddef} is what will be used to asses the future sensitivity at antimatter experiments. We will declare that a model is {\em within discrimination capabilities} of a given future experiment at X\% confidence level if it satisfies the relation
\begin{equation}
\chi^2\ >\ (\chi^2)^{X\%}_{n_{\rm b}}.
\end{equation}

We focus, for definiteness, on the case of the PAMELA detector, and compute the reduced $\chi^2$  for an effective area of 24.5 ${\rm cm}^2{\rm sr}$, an exposure time of 3 years, and resorting to a trial energy binning as sketched in ref.~\cite{pamelasens}\footnote{The data binning is not going to be homogeneous, since the fluxes have a peak at a few GeV energy. For definiteness, we divided the total logarithmic energy interval scanned by Pamela into three parts and estimated the respective number of bins from fig.~7 of ref.~\cite{pamelasens}.}. The results we will show are in the limit of known background, i.e. in the (optimistic) scenario in which degeneracies in the parameters used to model the propagation of charged cosmic rays in the Galaxy are resolved, say, by precision measurements of ratios of secondaries to primaries for several light cosmic-ray nuclei; in this context, we model the background according to the same estimates already implemented in the previous section. 

Fig.~\ref{fig:FuturePbarEplus} shows the future discrimination perspectives for the three benchmark models described in sec.~\ref{sec:SUSYmodels} for antiprotons (left) and positrons (right). The horizontal green line represents the 95\% C.L. exclusion limit for PAMELA, while the three lines indicate the quantity in the left-hand side of eq.~(\ref{eq:chisquareddef}), i.e. the projected $\chi^2$ for the given SUSY model. Notice that the two horizontal lines are slightly different, as the number of bins $n_{\rm b}$ for antiprotons is larger than that for positrons. 

For wino and higgsino-like lightest neutralinos, antiprotons will probe a larger portion of parameter space than in the case of positrons. On the other hand, a bino-like neutralino will give a signal which is not going to be detected even in the favorable case of the occurrence of an $s$-channel resonance.

We wish to make another step forward and present the results we just obtained in a form which is less closely tied to what it has been assumed as properties of a given detector. In fact, we definitely do not aim to assess in detail the detector responses, and what we extrapolated for PAMELA should be taken just as an approximation which can be improved with dedicated simulations. We wish instead to present a simple scheme to which an eventual refined analysis can refer to, without having to readdress from scratch the problem for the particle physics point of view; at the same time, this should be a valuable tool to make it easy to extend our analysis to other alternative particle physics scenario. With this goal in mind, we notice that the upcoming antimatter experiments PAMELA and AMS will collect a rather {\em large number} of positrons and antiprotons, and hence the number of energy {\em bins} in eq.~\ref{eq:chisquareddef} will be significantly large.  We find therefore appropriate to approximate the discrete sum in this equation with an integral. The region of integration will be given by the lowest and the largest energies, $T_{\rm min}$ and $T_{\rm max}$, accessible by the experiment. We henceforth define the quantity
\begin{equation}
I_\phi\equiv\ \int_{T_{\rm min}}^{T_{\rm max}} \frac{\left(\phi_s\right)^2}{\phi_b}{\rm d}E.
\end{equation}
The visibility condition, which reflects the continuum version of eq.~\ref{eq:chisquareddef}, is then given by
\begin{equation}\label{eq:IPHIdiscrim}
I_\phi\ >\ \frac{(\chi^2)^{X\%}_{n_{\rm b}}}{A\cdot T}.
\end{equation}
In the parameter $I_\phi$, the dependence on the extreme of integration is small as long as the peak on the signal to background ratio falls within them (and this is always the case for the models we are considering and for values expected for PAMELA and AMS);
we have assumed that the effective area is independent of energy, while, if this is not a good approximation, a weight function should be accordingly introduced. For the PAMELA setup already exploited, supposing a number of bins given by $n_b=55\div60$, and taking $(\chi^2)^{95\%}_{n_{\rm b}}\simeq 75$, we find the following critical values for $I_\phi$, after, respectively, one and three years of data taking:
\begin{equation}\label{eq:Iphis}
I_\phi^{1{\rm y}}=9.7\cdot 10^{-8}\, \, {\rm cm}^{-2}{\rm sr}^{-1}{\rm s}^{-1} \qquad I_\phi^{3{\rm y}}=3.2\cdot 10^{-8}\, \, {\rm cm}^{-2}{\rm sr}^{-1}{\rm s}^{-1}.
\end{equation}
We will refer to these values when comparing the perspectives of detection with different methods.

\FIGURE[t]{\epsfig{file=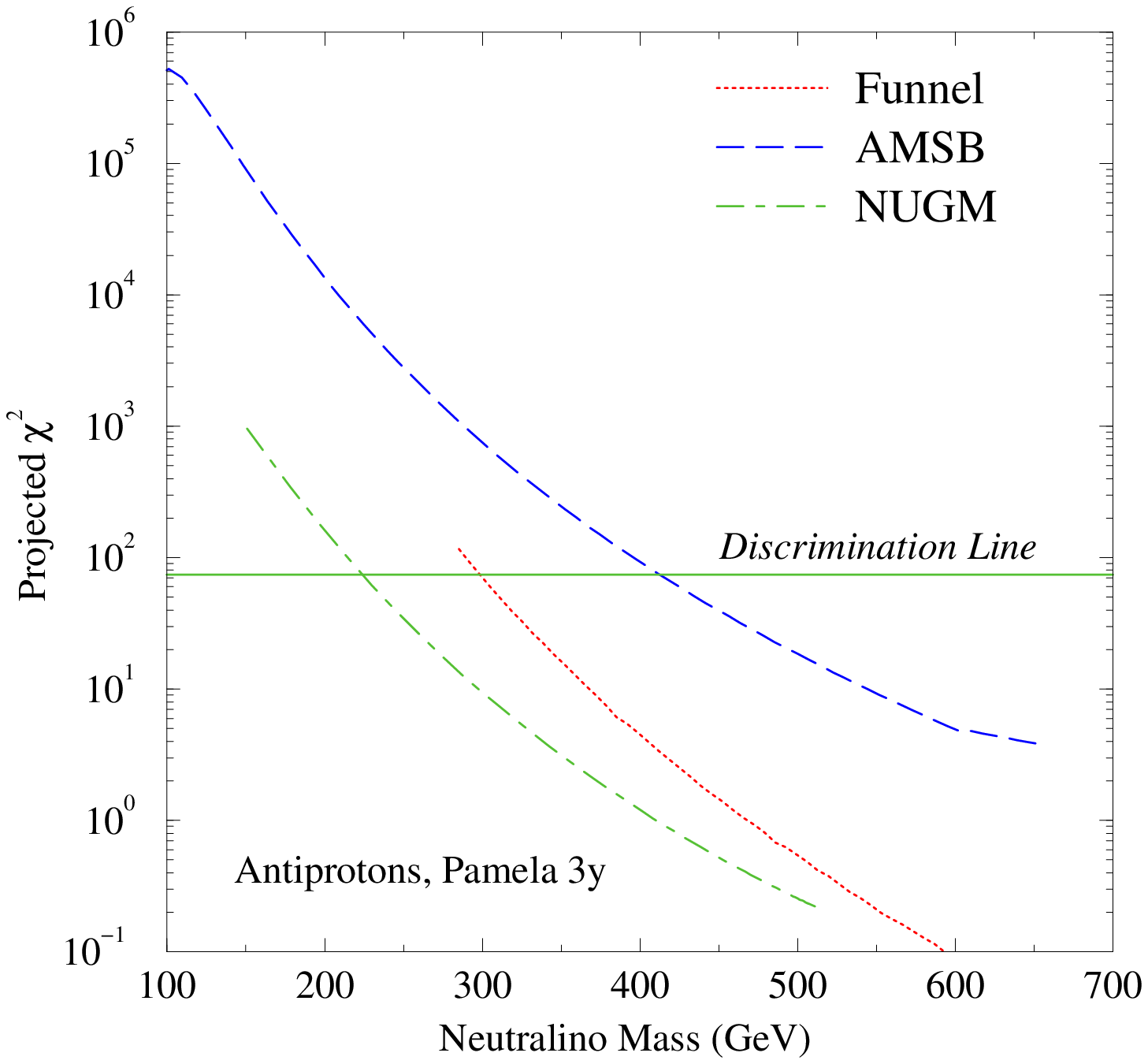,width=7.1cm}\quad
\epsfig{file=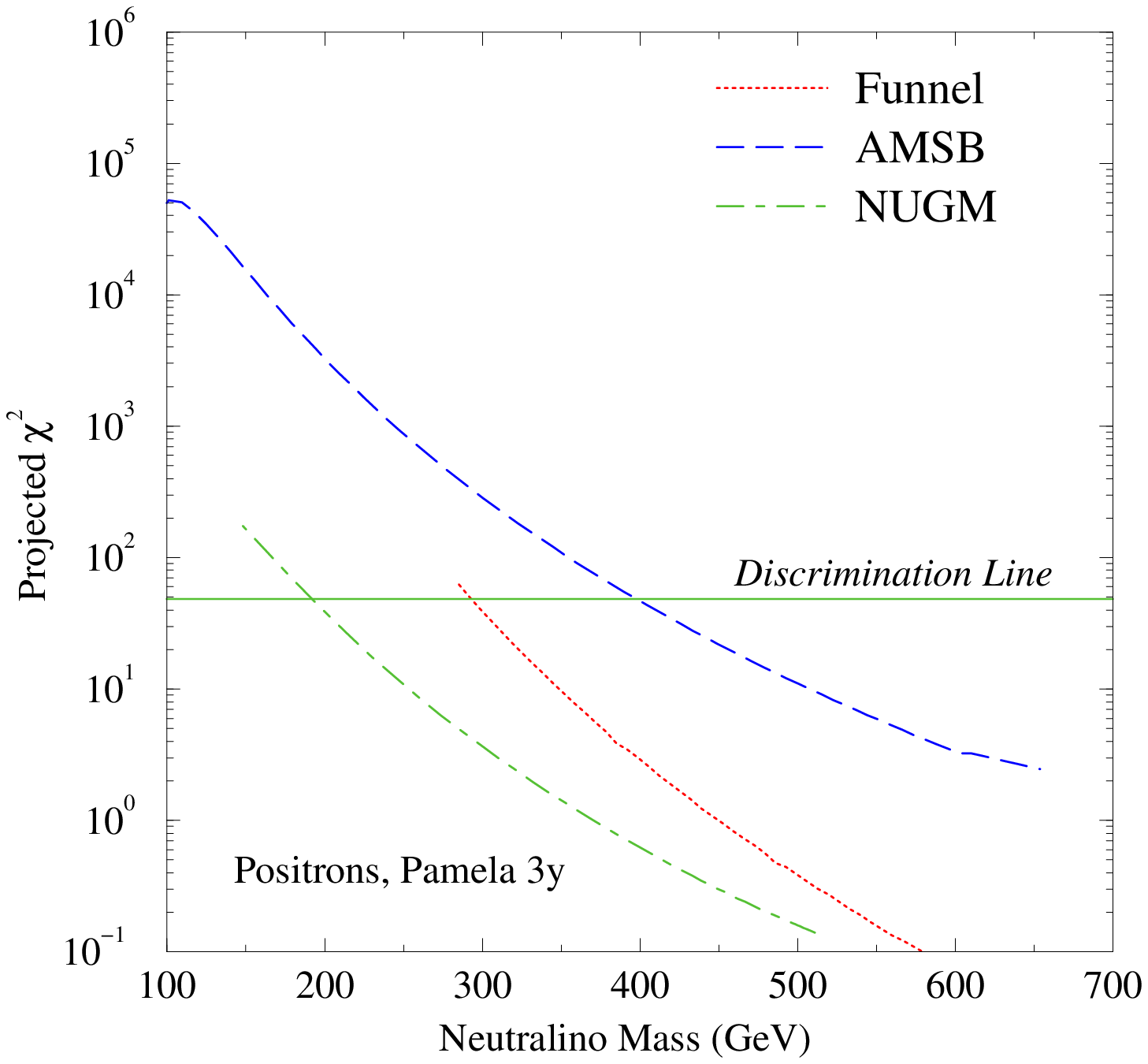,width=7.1cm} 
\caption{Discrimination perspectives on the three SUSY models of Tab.~\ref{tab:models} for the PAMELA experiment, as regards antiprotons (left) and positrons (right). The three lines correspond to the expected total $\chi^2$ for the three SUSY models, calculated assuming a small signal to background ratio (i.e. in the approximation of eq.~(\ref{eq:approximation})). The horizontal solid line reproduces the 95\% C.L. exclusion limit $\chi^2_{0.95}$ for the assumed PAMELA data binning.}
\label{fig:FuturePbarEplus}}

An ambitious goal for the future is to reach the experimental sensitivity to detect the antideuteron component in cosmic rays. The procedure we will follow to assess the detection perspectives of neutralino dark matter in this channel will be different from what we have just done for antiprotons and positrons. For antideuterons the background is expected to be totally negligible in the low energy regime~\cite{donato-pbar},
and even detection of 1 low-energy event would imply discovery of an exotic component. 
We refer, as the ultimate reach experiment in the future, to the gaseous antiparticle 
spectrometer (GAPS) \cite{GAPSproposal}. This is a proposal for an instrument
looking for antideuterons in the energy interval 0.1-0.4 GeV per nucleon, with
estimated sensitivity level of $2.6\times10^{-9}\textrm{m}^{-2}\textrm{sr}^{-1}\textrm{GeV}^{-1}\textrm{s}^{-1}$, to be placed either on a satellite orbiting
around the earth or on a probe to be sent into deep space. 
Fig.~\ref{fig:FutureDbar} shows the visibility ratios (i.e., the projected signal over projected sensitivity) for GAPS-like experiments placed on a satellite around Earth (left) and on a deep space probe (right). 

\FIGURE[t]{\epsfig{file=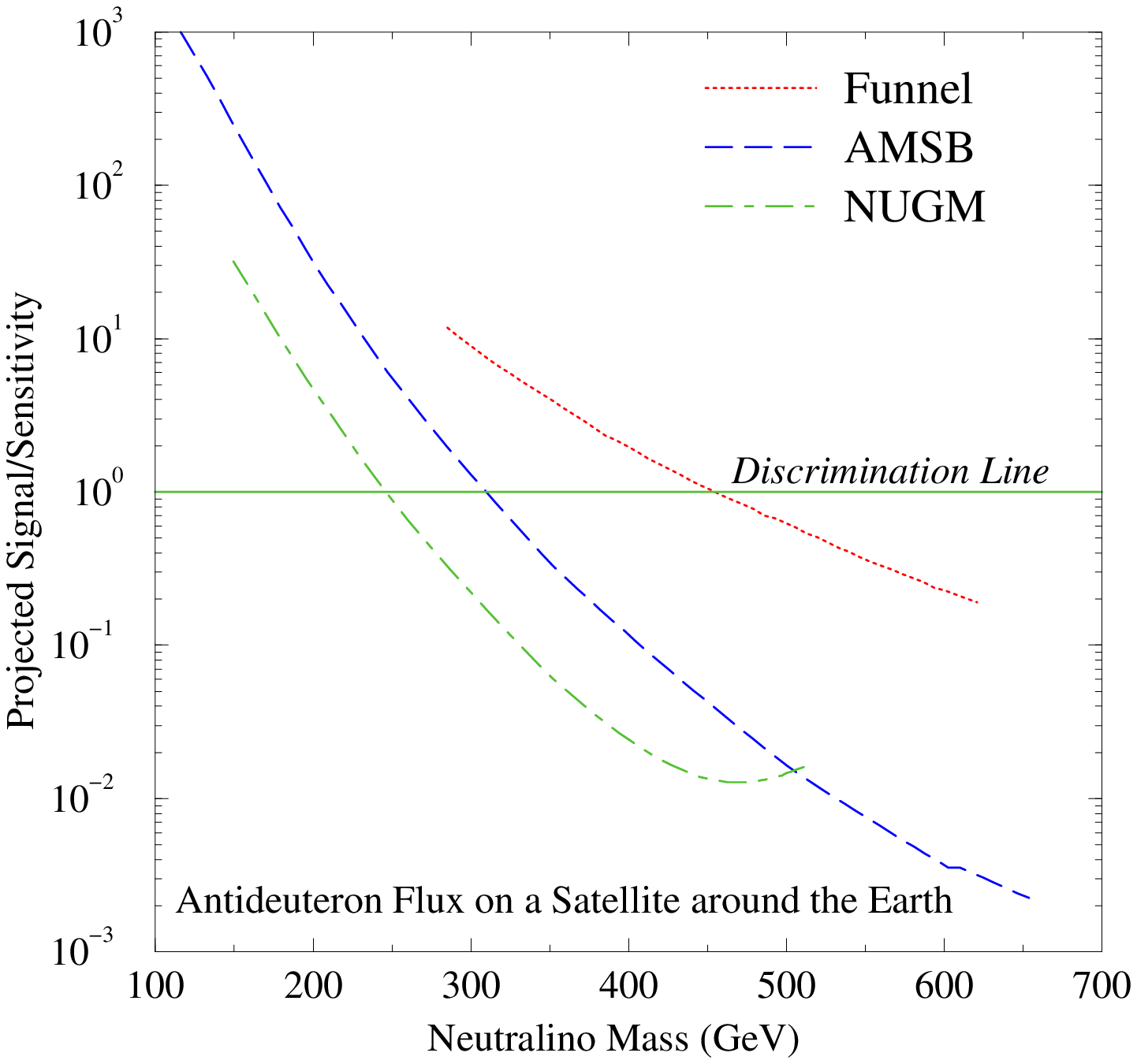,width=7.1cm}\quad
\epsfig{file=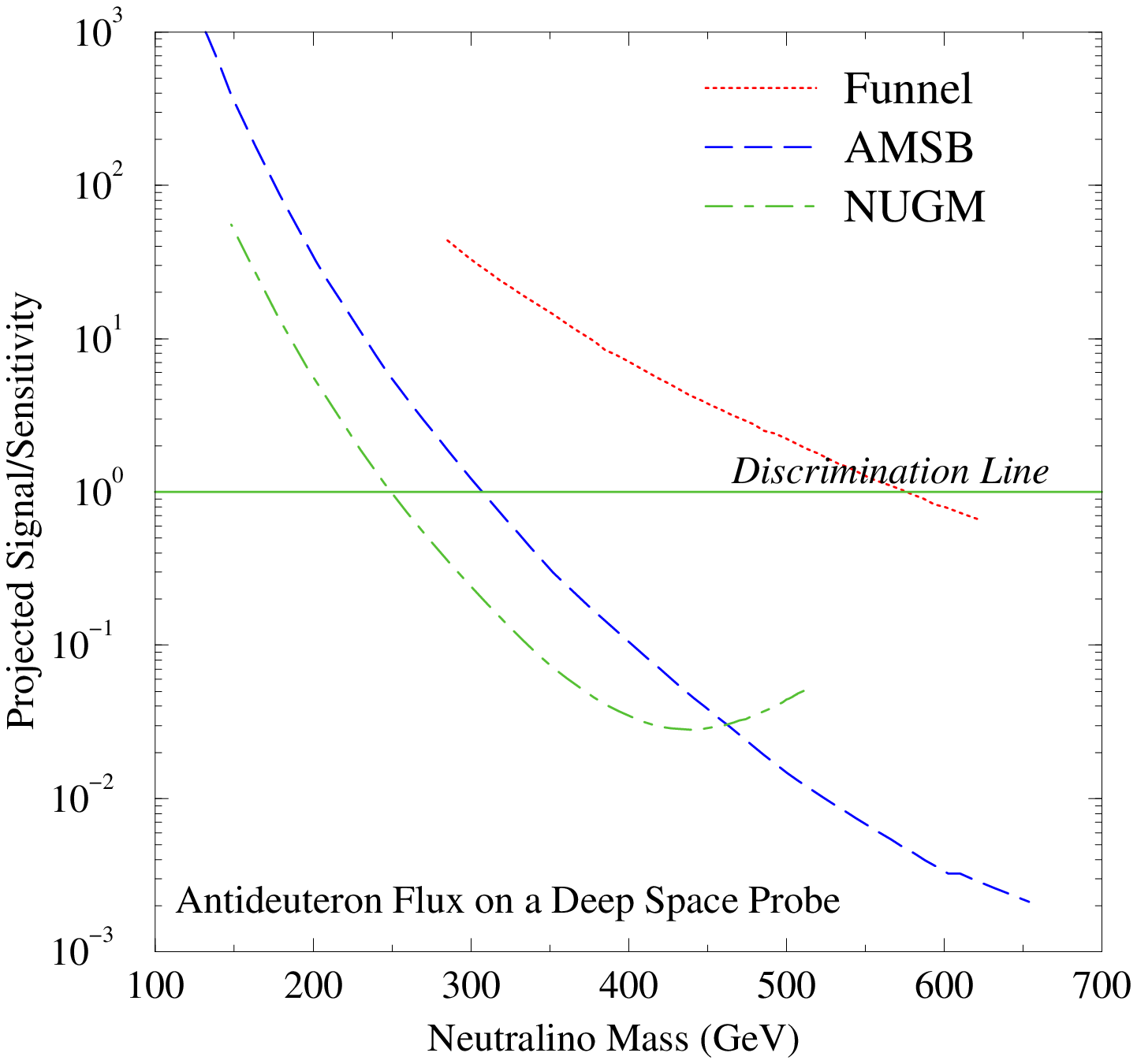,width=7.1cm} 
\caption{Discrimination perspectives on the three SUSY models of Tab.~\ref{tab:models} for a GAPS-like experiment for antideuteron searches in the antideuteron kinetic energy interval 0.1 GeV $<T_{\overline{D}}<$0.4 GeV, respectively for a satellite around Earth (right) and for a probe in deep space (left). On the $y$ axis we plot the expected signal-to-sensitivity ratio; the solid lines correspond to a signal from SUSY contributions which equals the assumed experimental sensitivity.}
\label{fig:FutureDbar}}

Remarkably, antideuterons searches in the low energy range will very efficiently probe the funnel region: the expected signal for the {\bf Funnel} benchmark model is in fact in this case more than one order of magnitude larger than that from winos or higgsinos. This enhancement of antideuteron production can be understood looking back at fig.~\ref{fig:dbIS}: antideuterons are efficiently produced by resonantly annihilating binos, which at large $\tan\beta$ mainly decay into $b\overline{b}$. Soft jets from $b$ decay then induce, by hadronization, the low energy tail of antiprotons and of antideuterons (see also fig.~\ref{fig:pbIS}). In the case of winos and higgsinos, instead, the main contribution comes from $W$ decays, and features a peak at larger energies, ranging from 1 to 10 GeV. At large enough neutralino masses, higgsino like neutralinos tend to produce soft antideuterons from $Z^0$ decays, as already pointed out from the left panel of fig.~\ref{fig:dbIS}, a feature which explains the anomalous rise in the visibility ratio for the {\bf NUGM} model in fig.~\ref{fig:FutureDbar}.


\subsection{Future search strategies and the role of antimatter}\label{sec:IPHI}

\FIGURE[t]{\epsfig{file=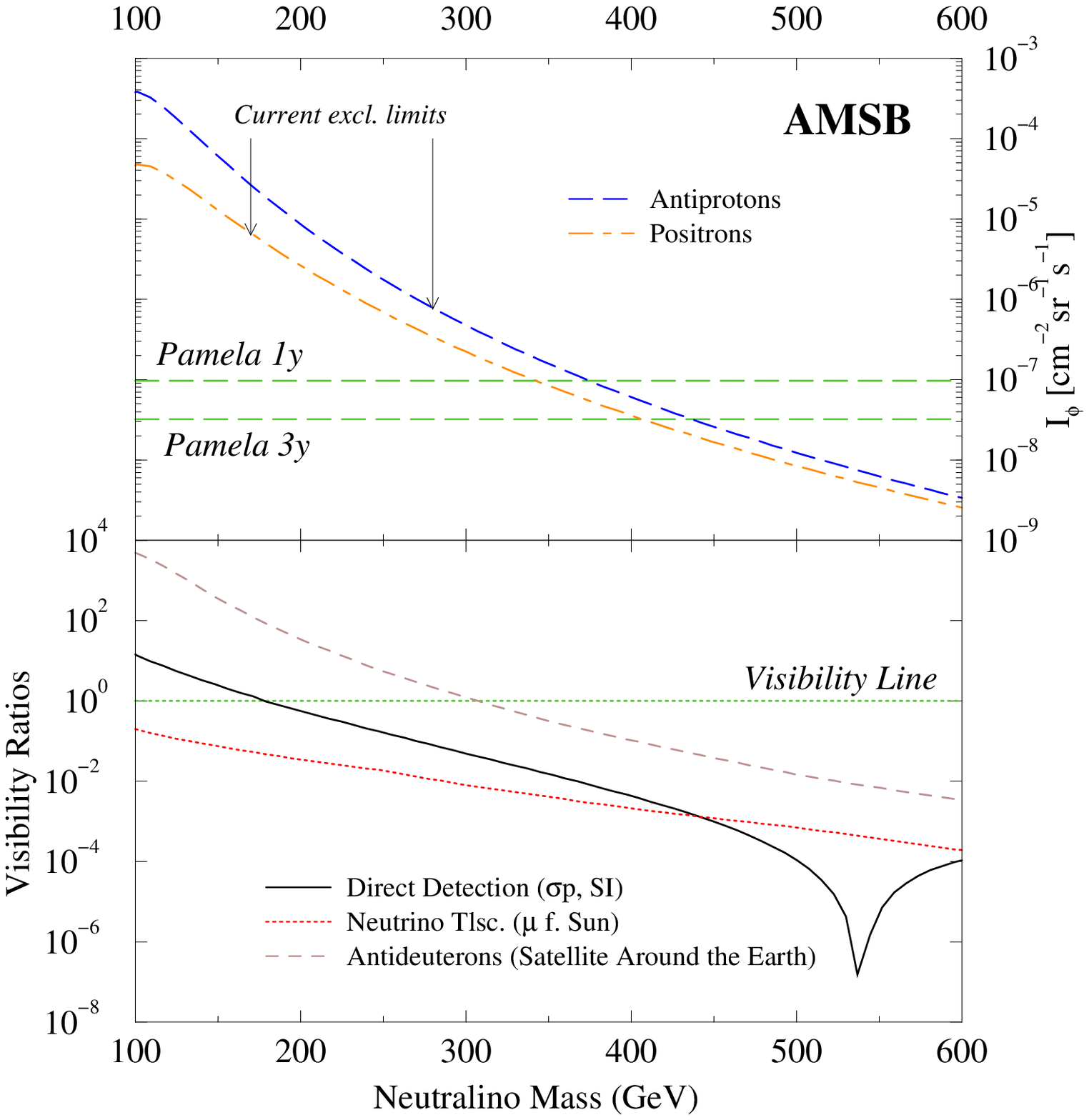,width=10.5cm}
\caption{Future exclusions limits on the {\bf AMSB} model from direct and indirect dark matter searches. In the upper panel we plot the discrimination parameter $I_\phi$ respectively for antiprotons (long dashed line) and positrons (dot-dashed), as well as the PAMELA experiment projected sensitivity after 1 and 3 years of data taking. The lower panel displays Visibility Ratios (expected signal over projected future exclusion limits) respectively for direct neutralino searches ($\sigma_{\chi-P}$, spin-independent) at XENON-1 ton \cite{Aprile:2002ef}, for indirect neutralino detection through the neutrino-induced muon flux detection from the center of the Sun at ICECUBE \cite{icecube}, and for antideuteron searches with a satellite around the Earth.}
\label{fig:FutureAMSB}}

In order to compare the sensitivity of future direct dark matter searches and searches with neutrino telescopes with that of antimatter, we will hereafter deal with the quantity $I_\phi$, rather than with the $\chi^2$, and plot the experimental discrimination sensitivity of PAMELA at different data taking times (see eq.~(\ref{eq:Iphis})). As regards, instead, antideuterons, we will keep on with the visibility ratio; we plot it together with the visibility ratio for direct spin-independent searches, assuming as reference sensitivity for future experiments that of the proposed XENON detector~\cite{Aprile:2002ef}, and for the muon-induced flux from the center of the Sun at neutrino telescopes, implementing the projected sensitivity of the km$^2$-size detector which is being built
by the ICECUBE Collaboration~\cite{icecube}.

Fig.~\ref{fig:FutureAMSB} compares future detection perspectives in the {\bf AMSB} benchmark model. Remarkably, the most promising detection strategies, as emerging from the upper panel of the figure, reside in antiproton searches, which in one year of data-taking will probe wino masses up to approximately 370 GeV. For comparison, direct detection at XENON will be able to detect masses not larger than 200 GeV, while this model will not give any signal at Neutrino Telescopes. Let us stress that what we find holds quite independently from the value of $m_0$. This in turns implies that for large values of $m_0$, where visibility at the LHC is going to be much suppressed, owing to a very heavy squark spectrum, antimatter searches will be the only way to probe this kind of anomaly mediated SUSY breaking models. We point out again that this conclusion holds in case these dark matter candidates are providing the bulk of the CDM in the galaxy (i.e. in presence of some mechanism to enhance the relic density), otherwise, rescaling fluxes, the discrimination capability is washed out.

\FIGURE[t]{\epsfig{file=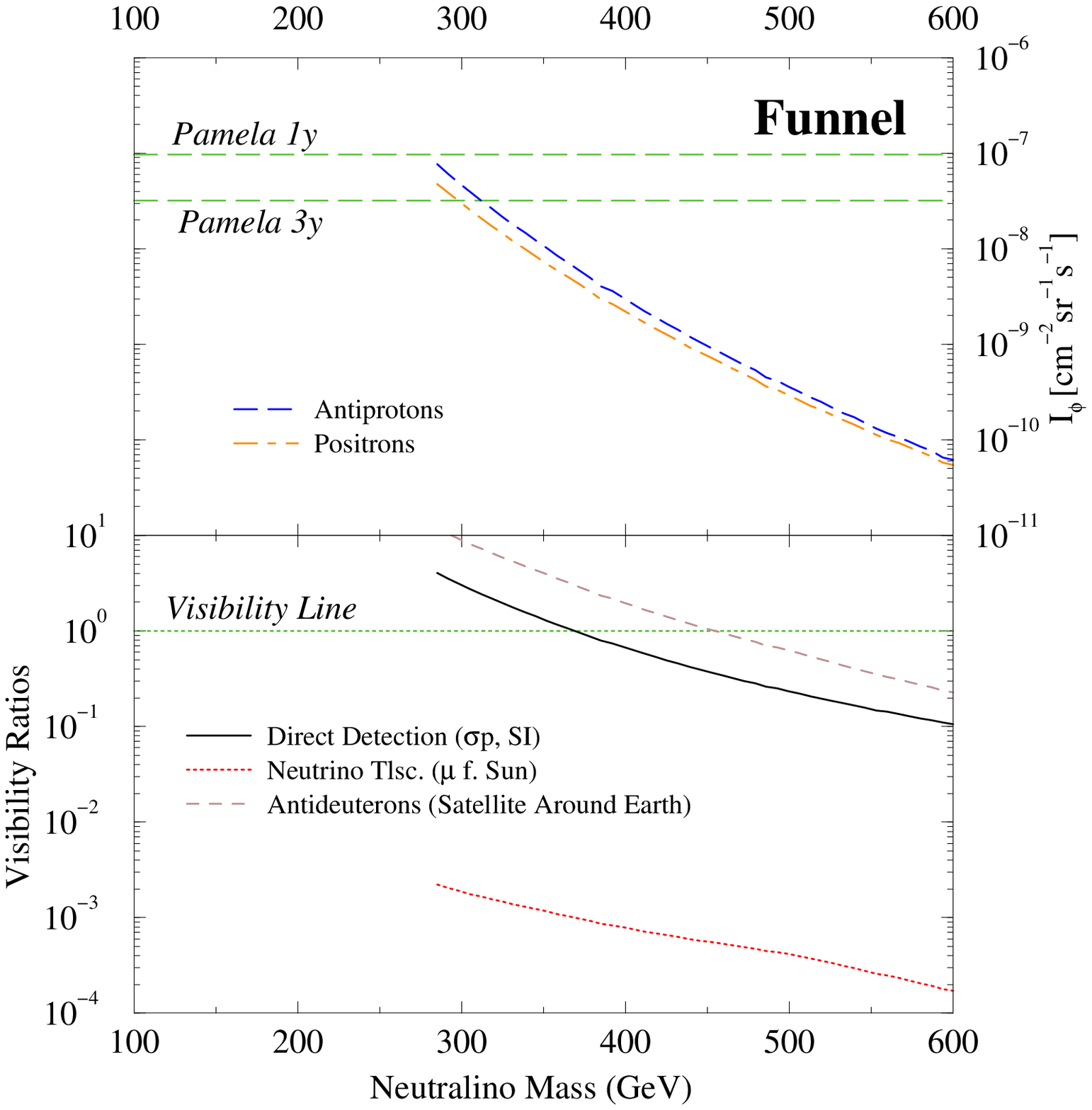,width=10.5cm}
\caption{Future exclusions limits on the {\bf Funnel} model. Symbols and conventions are the same as in fig.~\ref{fig:FutureAMSB}.}
\label{fig:FutureFUNN}}

As regards the {\bf Funnel} benchmark model, see fig.~\ref{fig:FutureFUNN}, antideuterons searches give the lion's share in detection perspectives, as previously pointed out. For this model, both antiprotons and positrons will not be able to probe much of the parameter space, since neutralino masses lie mostly above 300 GeV, but searches for low energy antideuterons might be able to reach significantly heavier masses, up to 450 GeV.  A comparison with standard direct detection strategies, whose discrimination capability is going to be less than 350 GeV, shows once again the utmost importance of antimatter searches in the quest for SUSY dark matter. 

Finally, fig.~\ref{fig:FutureNUGM} reproduces our results as far as the {\bf NUGM} benchmark model is concerned. In this case, antiprotons will reach a discrimination sensitivity of 250 GeV, and positrons of 200 GeV, after 3 years of Pamela data taking. For comparison, we see that antideuteron searches will do as good as antiprotons, while direct dark matter experiments will probably be marginally able to exclude models up to 400 GeV, though the visibility ratio lies in this case so close to 1 that this does not guarantee that XENON would see any signal, even at very low masses.

\FIGURE[t]{\epsfig{file=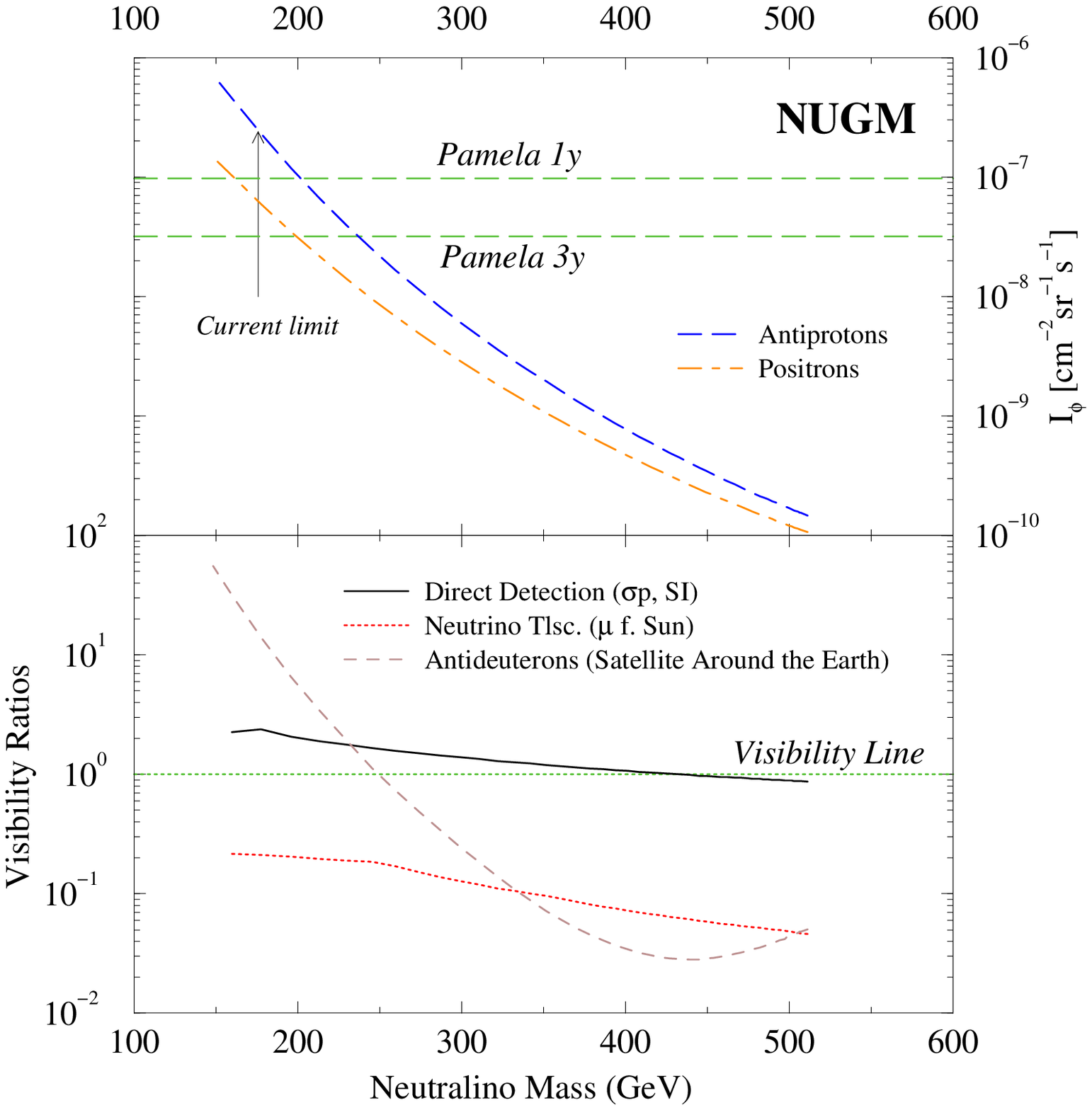,width=10.5cm} 
\caption{Future exclusions limits on the {\bf NUGM} model. Symbols and conventions are the same as in fig.~\ref{fig:FutureAMSB}.}
\label{fig:FutureNUGM}}


\section{The halo model dependence}\label{sec:hmodels}

All results presented so far are in the context of a rather conservative halo profile. The spherically symmetric Burkert model we have implemented, from the point of view of Galactic dynamics, gives a subdominant term in a Galaxy embedding a stellar disc close to maximal, i.e. a dark term which, in the inner portion of the Galaxy, is hardly providing any dynamical effect at all. From the point of view of structure formation, we have already mentioned that some mechanism has to be invoked to reconcile this final configuration with a large core with models with large cusp describing, according to the simulations, the Galaxy before the baryon infall. 

Of course, this is not the only consistent picture. In the regime in which the baryons settle in the inner portion of the Galaxy through a smooth and slow process, actually the back-reaction on the dark matter particle is expected to go in the opposite direction with respect to what we have assumed so far. This is the limit of adiabatic contraction of the system, with no net transfer of angular momentum between baryonic and non-baryonic terms; assuming spherical density profiles and unchanged local velocity distribution, the mass distributions in the initial and final configurations are related by~\cite{blumental}:  
\begin{equation}
M_i(r_i) r_i = \left[M_b(r_f)+ M_{CDM}(r_f)\right] r_f\;.
\label{eq:adia}
\end{equation}
Here, $M_i(r)$, $M_b(r)$ and $M_{CDM}(r)$ refer, respectively, to the mass profile of the halo before the baryon infall (i.e. the form one can infer from N-body simulation results), the baryon component as observed in the Galaxy today, and the cold dark matter component in its nowadays configuration, with the CDM cusp which is preserved or, actually, increased. It is possible to derive models for the Galaxy within this framework and fully consistent with dynamical measurements, see, e.g.,~\cite{KZS,halomod}. We consider here, as an example, the adiabatically contracted profile derived implementing the CDM profile found in ref.~\cite{n03} interpolating on the results from one of the simulations with highest resolution so far.  As the profile is derived numerically we cannot give its explicit form here; we just mention that it is obtained for a halo profile with virial mass and concentration parameter of, respectively, $M_{vir} = 1.8 \times 10^{12}\msun$ and $c_{vir} = 12$, and that the local density of the final spherically symmetric dark halo is equal to $\rho_{N03}(r_0)= 0.38$~GeV~cm$^{-3}$.
The approximation of adiabatic contraction is assumed to be valid down to the radius of 1~pc, below which it is assumed that process which as led to the formation of the central black hole in the Galaxy~\cite{bhobservation} has erased any eventual enhancement (i.e. we do not have a spike in the form predicted, e.g., in~\cite{gs}).  As for the previous halo model, this configuration as well has been found after implementing all available dynamical constraints, and by the corresponding velocity distribution is derived {\em self-consistently} (for details, see~\cite{halomod}).

\FIGURE[t]{\epsfig{file=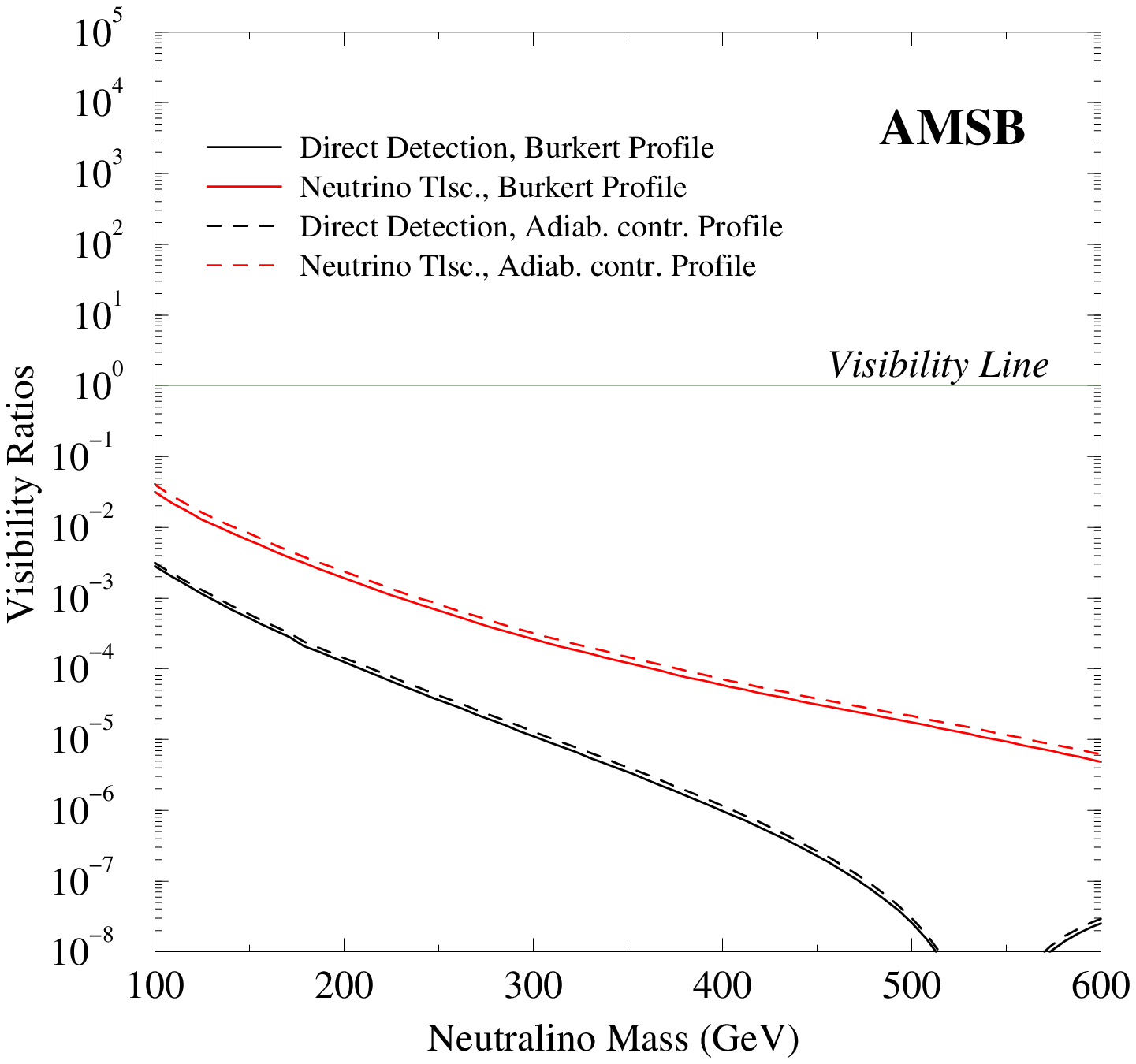,width=7.1cm}\quad
\epsfig{file=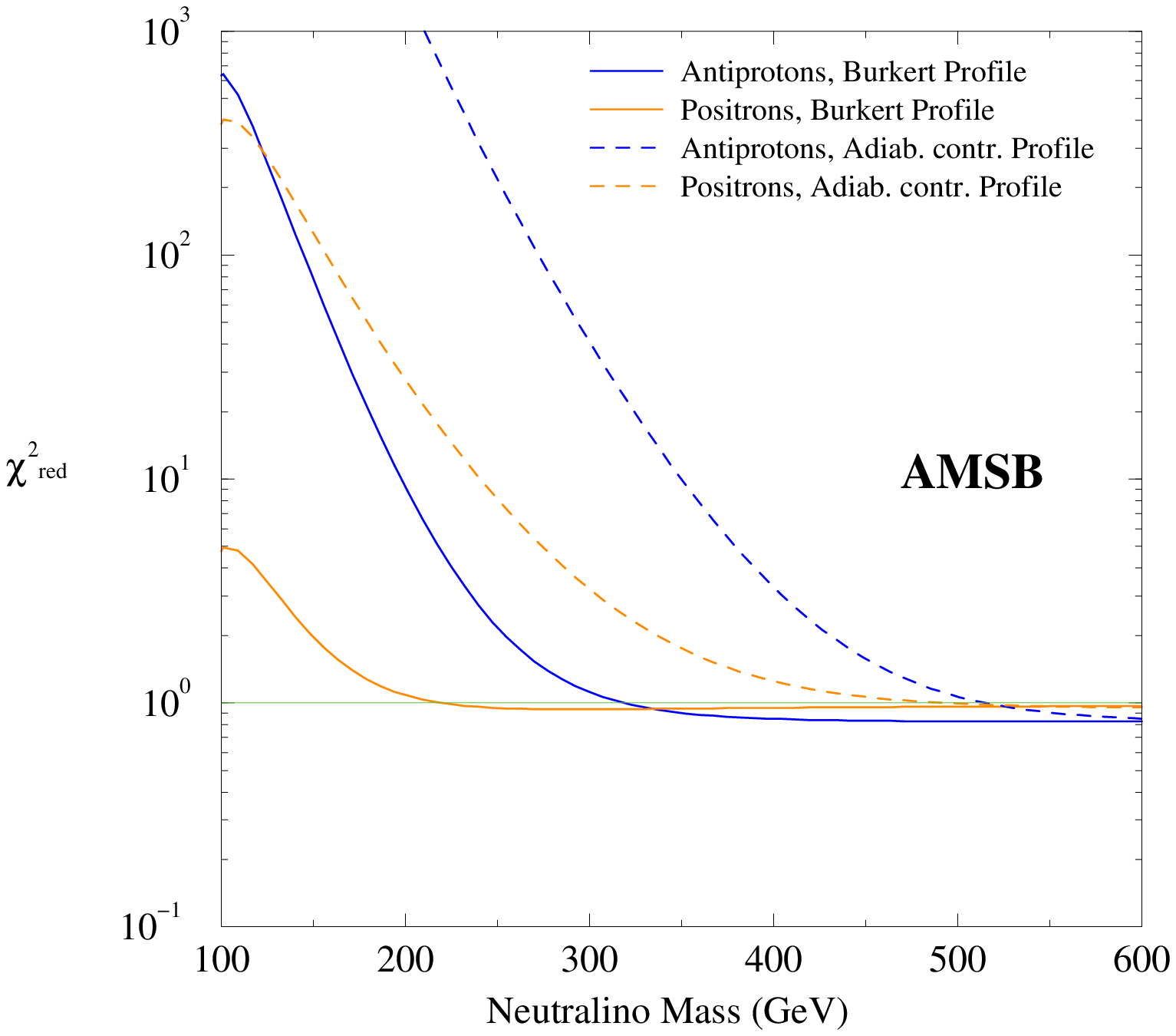,width=7.1cm} 
\caption{A comparison of the {\em current} dark matter exclusion limits, on the {\bf AMSB} model, for two different Halo Models, the {\bf Burkert} profile (solid lines) and the {\bf Adiabatically Contracted} profile (dashed lines). Right: direct ($\sigma_{\chi-P}$, spin independent) and indirect (muons from neutralino annihilations in the center of the Sun) detection Visibility Ratios. Left: the $\chi^2_{\rm red}$ for Antiprotons and Positrons.}
\label{fig:CurrentHM}}

\FIGURE[t]{\epsfig{file=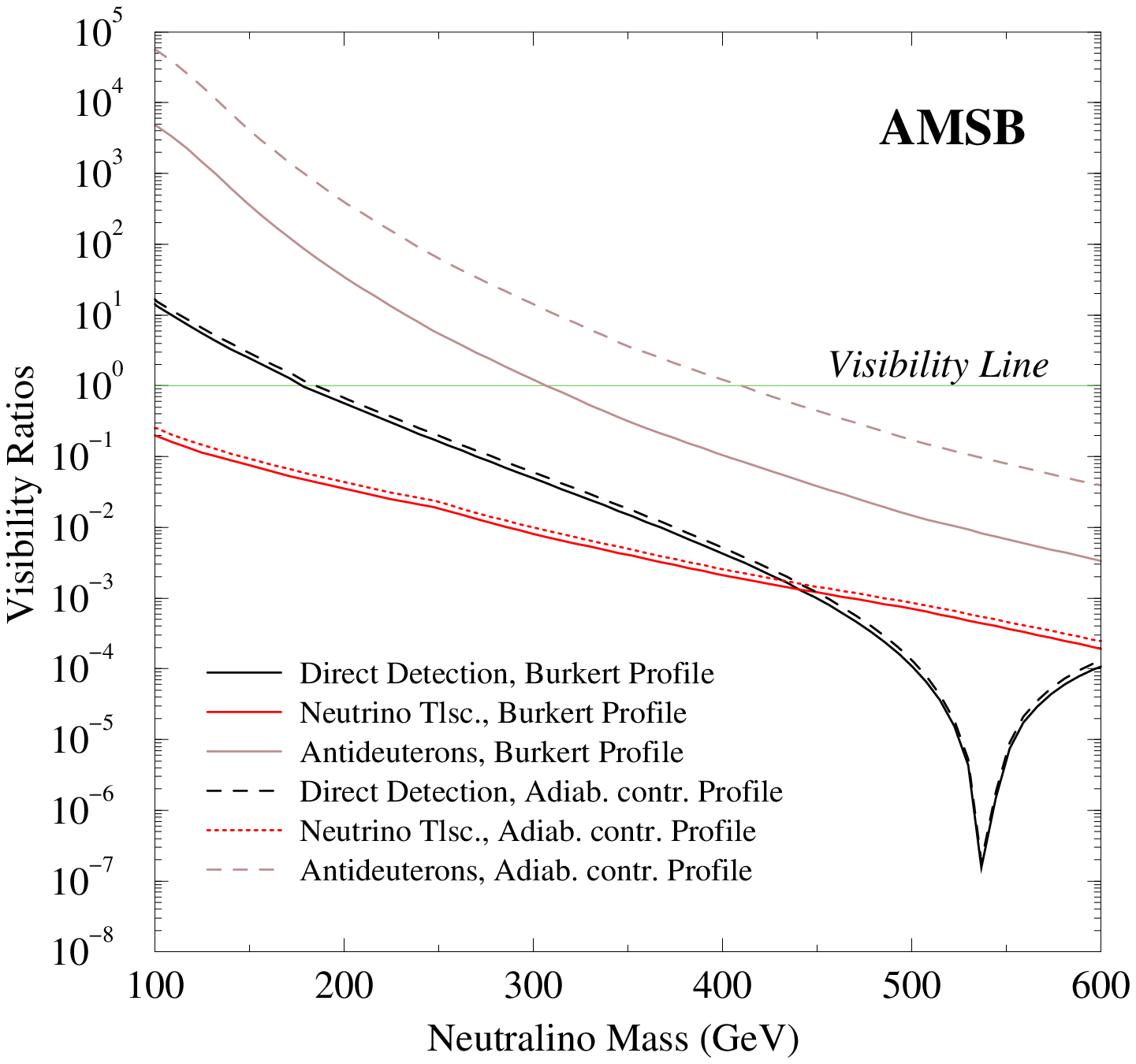,width=7.1cm}\quad
\epsfig{file=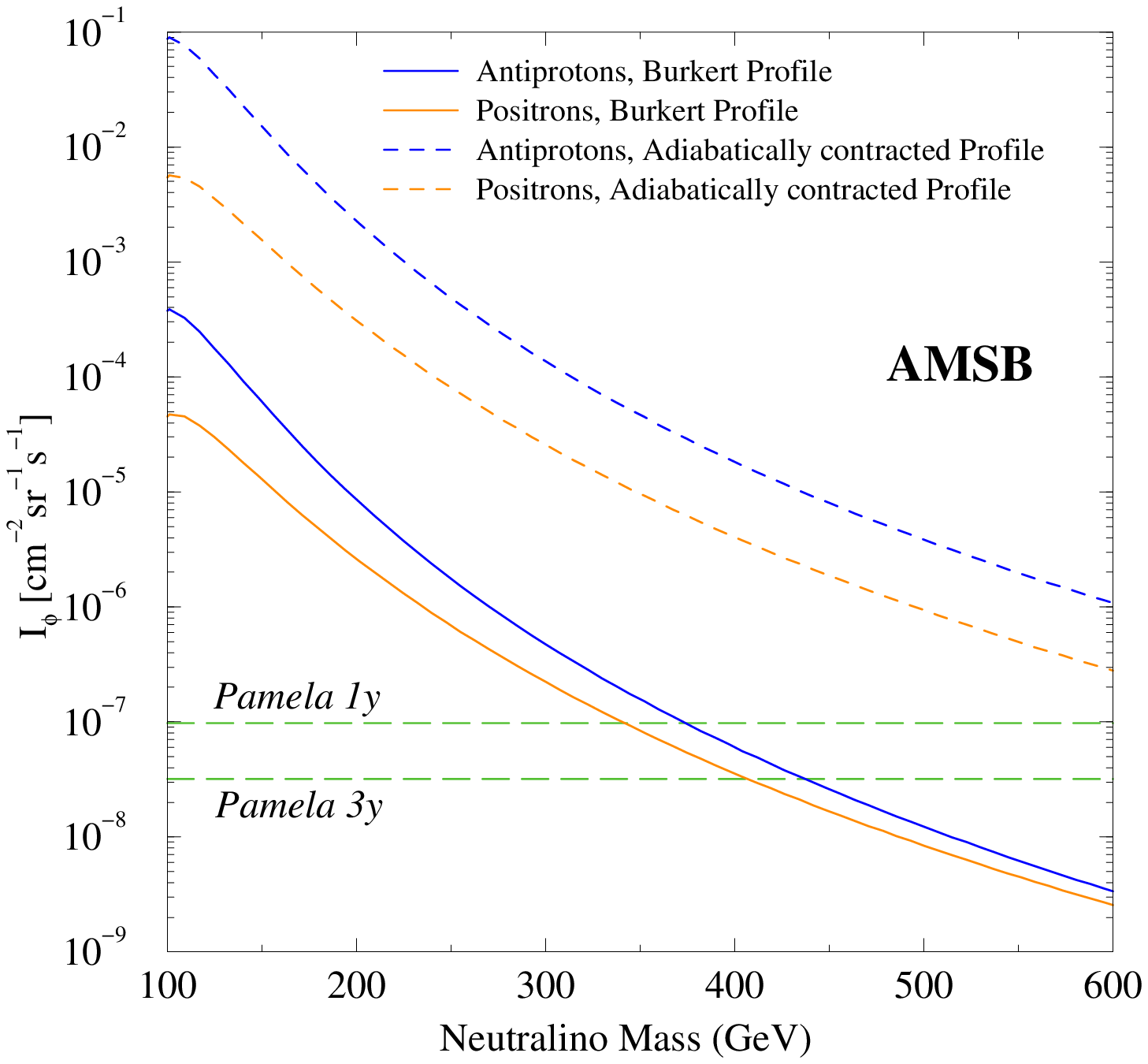,width=7.1cm} 
\caption{Future dark matter discrimination perspectives on the {\bf AMSB} model for the {\bf Burkert} (solid lines) and for the {\bf Adiabatically Contracted} (dashed lines) Halo Model profiles. The left panel shows the Visibility Ratios at XENON-1 ton direct detection and at ICECUBE, as in fig.~\ref{fig:CurrentHM}. We also indicate how discrimination perspectives from antideuteron searches on a satellite around Earth vary with the Halo Model. In the right panel we display instead the discrimination parameter $I_\phi$ for antiprotons and positrons, as well as the putative 1 and 3 years sensitivity of the PAMELA experiment (long dashed lines).}
\label{fig:FutureHM}}

Fig.~\ref{fig:CurrentHM} shows how the results for the current exclusion limits on the {\bf AMSB} benchmark model are affected resorting to the two different halo models we have introduced. Direct dark matter detection and neutrino telescopes rates are largely unaffected in the adiabatically contracted profile, and the conclusions we drew for the Burkert profile still apply. Antimatter searches are instead largely boosted by the new profile, which features a larger dark matter matter densities towards the Galactic center: antiproton flux fits rule out, {\em with current data} models with neutralino masses as large as 470 GeV, and positrons up to 370 GeV. Turning to future perspectives, fig.~\ref{fig:FutureHM} illustrates that, again, the situation is not dramatically different for direct detection and neutrino telescopes, as compared with what shown in fig.~\ref{fig:FutureAMSB}; on the other hand, while antideuterons would probe masses 100 GeV larger than in the Burkert profile, strikingly enough we find that {\em both antiproton and positron searches will probe, in even only 1 year of data taking, the whole parameter space at the PAMELA experiment!}


\FIGURE[t]{\centerline{\epsfig{file=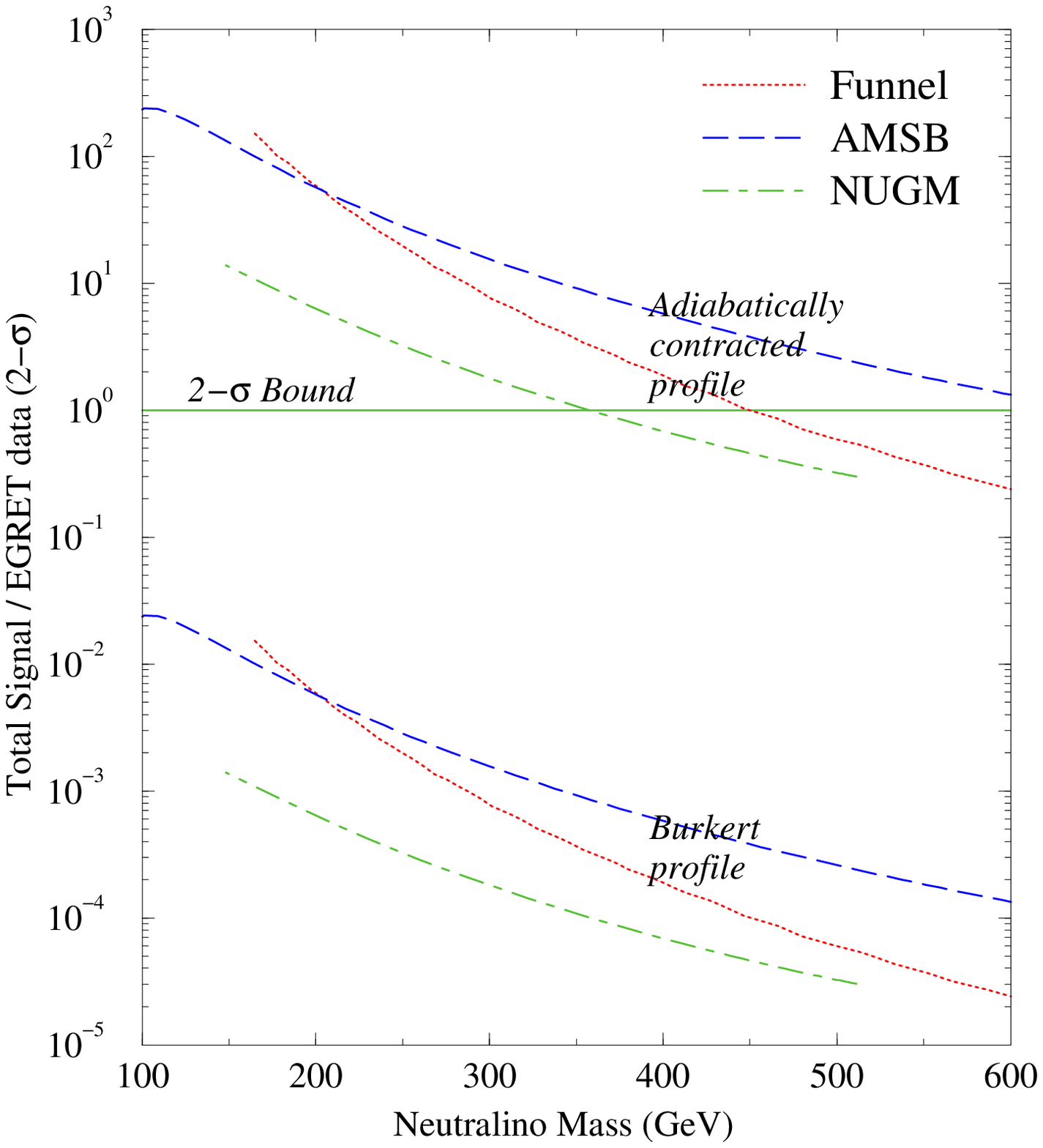,width=8.5cm}} 
\centerline{\caption{The bound coming from the continuous gamma rays background as measured by EGRET in the bin with the largest sampled energies ($4\, {\rm GeV} < E_\gamma < 10\, {\rm GeV}$), for the two halo models considered here and for the three benchmark scenarios. As a function of the Neutralino mass, we plot the ratio of the expected signal in the considered energy bin over the 2-$\sigma$ upper bound from the actual measured datum. Models above the horizontal green line are not consistent with EGRET and the particular considered halo model, and are therefore ruled out. The upper lines represent the results for the adiabatically contracted halo profile, while the three lower lines those for the Burkert profile.}
\label{fig:Gammas}}}

Since the adiabatically contracted profile is cuspy towards the galactic center, we expect that another indirect detection channel, namely the observation of gamma rays in the direction of the center of the Galaxy, could also probe a large fraction of the {\bf AMSB} model parameter space. The EGRET experiment, on the Compton Gamma-Ray Observatory has resolved a $\gamma$-ray source towards the GC~\cite{MH}, tentatively extended ($\sim 1.5^{\circ}$, of the order of the EGRET angular resolution) rather than point-like, and with a spectrum apparently incompatible with the spectrum expected for the diffuse $\gamma$-ray flux due to  the interaction of primary cosmic rays with the interstellar medium, while possibly consistent with a WIMP-induced component~\cite{cesarini}. Expected fluxes within our frameworks are compared
with the intensity and spectrum of the EGRET $\gamma$-ray source; the reader should also keep in mind that alternative explanations for this source have been proposed~\cite{pohl}, as well as that it has been argued that the position of the source should not be identified with the Galactic center~\cite{hooper}. A clearer statement on both these points will be feasible with the upcoming measurements by the next gamma-ray mission in space, the GLAST satellite \cite{glast}, and, eventually, it will be possible to derive even more stringent bounds.

As shown in fig.~\ref{fig:Gammas}, it turns out that the adiabatically contracted halo profile is greatly constrained by current data from gamma rays, which excludes, at the 2-$\sigma$ level, masses up to 600 GeV within the {\bf AMSB} benchmark model, as well as puts significant constraints for the other frameworks. It should be noticed however that the signal in gamma-rays is totally dominated by contributions close to the Galactic center rather than evenly distributed along the line of sight, hence the signal is extremely sensitive to what has been assumed on the (essentially unconstrained) distribution of dark matter in the Galactic center region; in that respect, note that for the signal for the Burkert profile is suppressed by over four orders of magnitude. Antimatter fluxes are much less dependent on this specific feature, and halo model configurations implying an enhancement in the antimatter fluxes without overproducing gamma-rays are certainly viable.

Finally, there are further effects inducing large enhancements for antimatter fluxes and affecting less critically the other signals. E.g., one can consider halo models with are flattened towards the Galactic plane, rather than being spherical: an increase in the local halo density enters linearly in direct detection and neutrino telescope rates, while quadratically for cosmic ray fluxes (the effect of flattening has been sketched, e.g., in~\cite{bottino-pbar}). Moreover, the dark matter halo may not be perfectly smooth but have a clumped component, again with possibly large enhancements in antimatter yields, see, e.g.,~\cite{clumpy}.

\section{Conclusions}\label{sec:conclusions}

We considered in this paper three benchmark scenarios, respectively featuring a bino, wino and a higgsino-like lightest neutralino. All models have large neutralino pair-annihilation cross sections, and the resulting relic abundance is compatible with the WMAP range provided some cosmological relic density enhancement mechanism, or non thermal neutralino production, is operating. We described in detail the spectral features of antiprotons, positrons and antideuterons for each of these models. 

Making use of a self-consistent halo model, we first directly compared the current performance of different search strategies, namely direct spin-independent, neutrino flux from the Sun and indirect detection through positron and antiproton fluxes; we concluded that, in all cases, {\em the only constraints presently coming from dark matter searches on the viable parameter space of the models under consideration stem from antimatter searches}. In many instances, and although available experimental data on antimatter have a poor statistics, the efficiency of antimatter searches currently does better than direct detection by various orders of magnitude.

Regarding perspectives for dark matter searches at forthcoming experimental facilities, we introduced a new parameter $I_\phi$ which allows, given a SUSY model, to reliably assess its visibility at future experiments, without any bias from the kind of data analysis which will be carried out. In particular, we provided the expected critical values of $I_\phi$ for the discrimination capability of the PAMELA experiment after one and three years of data taking. 

The comparison of future experimental dark matter search strategies in the three benchmark models considered  once again shows that antimatter searches may be highly competitive with respect to both direct detection and neutrino telescopes. In some cases, such as for antiprotons in a wino dark matter scenario or for antideuterons within a mSUGRA funnel scenario, antimatter searches may be {\em the only viable dark matter detection technique}.

Finally, we showed how our conclusions would be affected taking into account a more {\em cuspy} profile: while direct detection and neutrino telescope rates are hardly affected, antimatter fluxes are largely enhanced. The typical increase in the value of the $I_\phi$ parameter is between two to three orders of magnitude. Although the particular halo profile we considered features large gamma rays yields, different models for the Dark Matter distribution at the very center of the Galaxy would suppress this signal, while the expected antimatter fluxes would realistically lie between the two extreme cases we discussed here.

In the context of SUSY models with large annihilation cross sections, and in view of the imminent launch of space-based dedicated experiments, antimatter searches are therefore to be considered as a highly promising path towards the detection of dark matter.

\acknowledgments{P.U.\ was supported in part by the RTN project under 
grant HPRN-CT-2000-00152 and by the Italian INFN under the
project ``Fisica Astroparticellare''.}

\end{document}